\documentclass[useAMS,usenatbib]{mn2e}
\usepackage{graphicx}
\usepackage{natbib}
\usepackage{url}
\newcommand{\zab}{\ensuremath{z_\textrm{\scriptsize abs}}}
\newcommand{\kms}{\,km\,s$^{-1}$}
\newcommand{\jms}{J. Mol.\ Spectrosc.}

\newcommand{\soviet}{Sov. J. Exp. Theor. Phys.}
\newcommand{\ieee}{IEEE Trans. Aut. Contr.}
\newcommand{\molphys}{Mol. Phys.}
\newcommand{\astrnachr}{Astron. Nachr.}

\title[Constraint on a varying $\mu$ from H$_{2}$ in J1237+0647]{Constraint on a varying proton-to-electron mass ratio from molecular hydrogen absorption toward quasar \mbox{SDSS J123714.60+064759.5}}

\author[Dapr\`a et al.]{M. Dapr\`a,$^{1}$ J. Bagdonaite,$^{1}$  M. T. Murphy,$^{2}$ and W. Ubachs,$^{1}$\\
$^1$Department of Physics and Astronomy, LaserLaB, VU University,
  De Boelelaan 1081, 1081 HV Amsterdam, The Netherlands\\
$^{2}$Centre for Astrophysics and Supercomputing, Swinburne University of Technology, Melbourne, Victoria 3122, Australia\\
}

\begin{document}

\date{}

\pagerange{\pageref{firstpage}--\pageref{lastpage}} \pubyear{2015}

\maketitle

\label{firstpage}

\begin{abstract}
Molecular hydrogen transitions in the sub-damped Lyman $\alpha$ absorber at redshift $\zab\simeq2.69$, toward the background quasar \mbox{SDSS J123714.60+064759.5}, were analyzed in order to search for a possible variation of the proton-to-electron mass ratio $\mu$ over a cosmological time-scale. The system is composed of three absorbing clouds where 137 H$_{2}$ and HD absorption features were detected. The observations were taken with the Very Large Telescope/Ultraviolet and Visual Echelle Spectrograph with a signal-to-noise ratio of \mbox{32 per 2.5\kms pixel}, covering the wavelengths from 356.6 to 409.5 nm. A comprehensive fitting method was used to fit all the absorption features at once. Systematic effects of distortions to the wavelength calibrations were analyzed in detail from measurements of asteroid and `solar twin' spectra, and were corrected for. The final constraint on the relative variation in $\mu$ between the absorber and the current laboratory value is \mbox{$\Delta\mu/\mu = (-5.4 \pm 6.3_{\textrm{\small{stat}}} \pm 4.0_{\textrm{\small{syst}}}) \times 10^{-6}$}, consistent with no variation over a look-back time of 11.4 Gyrs.
\end{abstract}


\section{Introduction}
\label{sec:intro}
The investigation of highly redshifted absorption systems in the line-of-sight of quasars, by means of high-resolution spectroscopic observations, has become an established and powerful method to constrain a possible variation of the laws of physics over cosmological time. The variation of dimensionless fundamental constants, appearing as key building blocks in the laws of physics, is targeted in a comparison between astrophysical observation with laboratory measurement. For example, the fine-structure constant $\alpha=e^{2}/(4\pi\epsilon_{0}\hbar c)$ is probed via metal absorption searching for temporal \mbox{\citep{Webb1999}} and spatial variations \mbox{\citep{Webb2011}}.

The other dimensionless constant of nature determining the structure of molecular matter, the proton-to-electron mass ratio \mbox{$\mu \equiv m_{p}/m_{e}$}, may be targeted by observing absorption lines of a wide variety of molecules \mbox{\citep{Jansen2014}}. While radio astronomical observations of ammonia~\citep{Murphy2008, Kanekar2011} and of methanol molecules~\citep{Bagdonaite2013a,Bagdonaite2013b,Kanekar2015} have proven to be very sensitive, constraining $|\Delta\mu/\mu|$ at the level of \mbox{$\sim 10^{-7}$}, there are are only two radio sources at extragalactic distances where these molecules are observed, and these lie at redshifts $z<1$.

Molecular hydrogen (H$_{2}$) is a target to investigate a possible variation of $\mu$ at higher redshifts, where it is  observed in larger numbers of absorption systems. H$_{2}$ has many spectral lines that are sensitive to a variation in $\mu$~\citep{Thompson1975,Varshalovich1993,Ubachs2007}. Molecular hydrogen absorbing galaxies often display up to 100 H$_{2}$ spectral lines, which helps in improving the statistical basis of measurements made with this technique. Most importantly, H$_{2}$ absorbing systems can be observed, using ground-based telescopes, at redshifts $z>2$, even extending to $z=4.22$~\citep{Bagdonaite2015}, where the Lyman and Werner bands are redshifted into the optical band.

So far eight H$_{2}$ absorbing systems at $z>2$ have been analyzed for $\mu$-variation, most notably and most accurately the \mbox{J2123--005} system at \mbox{\emph{z}=2.05}~\citep{Malec2010,Weerdenburg2011}, the \mbox{HE0027--1836}~\citep{Rahmani2013} and the \mbox{Q2348--011} system~\citep{Bagdonaite2012} both at \mbox{\emph{z}=2.42}, the \mbox{Q0405--443} system at \mbox{\emph{z}=2.59}~\citep{Reinhold2006,King2008,Thompson2009}, the \mbox{Q0642--504} system at \mbox{\emph{z}=2.66}~\citep{Bagdonaite2014,Vasquez2014}, the \mbox{Q0528--250} system at \mbox{\emph{z}=2.81}~\citep{Reinhold2006,King2008,King2011}, the \mbox{Q0347--383} system at \mbox{\emph{z}=3.02}~\citep{Reinhold2006,King2008,Thompson2009,Wendt2011,Wendt2012}, and the \mbox{Q1443+272} system at \mbox{\emph{z}=4.22}~\citep{Bagdonaite2015}. A general conclusion from these studies is that the proton-to-electron mass ratio is constrained to \mbox{$|\Delta\mu/\mu| < 10^{-5}$} for redshifts in the range \mbox{\emph{z}=2-4.2}, corresponding to look-back times of up to 90 per cent of the age of the Universe. A review of the results of the analyses of H$_{2}$ absorbing systems has been given by \cite{Ubachs2011} and by \cite{Wendt2014}.

In the present study H$_{2}$ absorption at \emph{z}=2.69 is investigated in the line of sight toward quasar \mbox{SDSS J123714.60+064759.5}, hereafter \mbox{J1237+0647}, for further constraining a variation of the proton-to-electron mass ratio. The importance of this system is that carbon monoxide (CO) absorption in the optical domain is observed alongside with H$_2$~\citep{Noterdaeme2010}. This may be the basis for a simultaneous $\mu$-variation analysis using H$_2$ and CO~\citep{Salumbides2012}. As a first step the H$_2$ absorption in this system is analyzed for constraining $\Delta\mu/\mu$, results of which are presented in this paper, and this will serve as the basis for an analysis of the CO lines in a later paper.

\section{Data}
\label{sec:data}
The analysis presented in this work is based on observations made under four different observing programs, all carried out using the Ultraviolet and Visual Echelle Spectrograph (UVES) mounted on the 8.2m Very Large Telescope (VLT) at Paranal, Chile \citep{Dekker2000}. UVES is a cross-dispersed echelle spectrograph with two arms that are functionally identical: one covers the wavelengths in the range \mbox{300--500 nm} (Blue) and the other covers the range \mbox{420--1100 nm} (Red). Two dichroic beam splitters can be used to work in parallel with the two arms, centering them at different wavelengths. The details of the observational campaigns are presented in Table \ref{dataset}.

The raw 2D data were reduced following the procedure used by \cite{Bagdonaite2014}. The Common Pipeline Language version of the UVES pipeline was used to bias correct and flat field the exposures and then to extract the flux.  The ThAr lamp exposures were used for wavelength calibration; the ThAr flux was extracted  by using the same object profile weights as in a corresponding quasar exposure. After the standard reduction, the custom software \textsc{UVES\_popler}\footnote{\url{http://astronomy.swin.edu.au/~mmurphy/UVES_popler/}} was used to combine the extracted echelle orders into a single 1D spectrum which was then manually inspected and cleaned from bad pixels and other spectral artifacts, and the continuum was fitted with low-order polynomials.

The `final spectrum', obtained by combining all the exposures together, covers the wavelengths from 329.0 to \mbox{960.0 nm} with a signal-to-noise ratio (S/N) of \mbox{26 per 2.5\kms per pixel} at \mbox{$\sim$400 nm} in the continuum.

\begin{table*}
\centering
\caption{Details of the J1237+0647 quasar observations with UVES/VLT included in the present analysis. The total integration time, summed over 17 exposures, is 19.9 hrs. Most of the listed exposures were followed by attached ThAr calibrations and additional `supercalibrations' as indicated (see Section \ref{sec:supercali}). The rest of the data were calibrated using the regular ThAr exposures taken at the end of the night. The CCD binning was \mbox{2$\times$2} and the slit width was {1.0 arcsec} for all frames. Data collected under programs 082.A-0544(A) and 083.A-0454(A) were retrieved from the ESO archive.}
\label{dataset}
\begin{tabular}{*{6}{c}}
\hline
Program ID & Date & Integration time [s] & Central $\lambda$ [nm] & ThAr & Supercalibration\\
\hline
\hline
082.A-0544(A) & 27-03-2009 & 5400 & 390 + 564 & Regular & No\\
 & 27-03-2009 & 5400 &  & Regular & No\\
	& 29-03-2009 & 5400 & & Regular & No\\
 & 29-03-2009 & 5400 & & Regular & No\\
083.A-0454(A) & 27-04-2009 & 4500 & 390 + 775 & Regular & No\\
 & 27-04-2009 & 4500 & & Regular & No\\
091.A-0124(A) & 15-05-2013 & 4800 & 390 + 580 & Attached & Yes\\
 & 15-05-2013 & 1727 & & Attached & No\\
 & 15-05-2013 & 4800 & & Attached & Yes\\
 & 16-05-2013 & 4800 & & Attached & Yes\\
 & 16-05-2013 & 2025 & & Attached & No\\
093.A-0373(A) & 23-03-2014 & 4800 & 390 + 580 & Attached & Yes\\
 & 03-04-2014 & 4800 & & Attached & Yes\\
 & 28-05-2014 & 4800 & & Attached & Yes\\
 & 29-05-2014 &  826 & & Attached & No\\
 & 31-05-2014 & 4800 & & Attached & Yes\\
 & 02-06-2014 & 2800 & & Attached & No\\
\hline
\end{tabular}
\end{table*}

\subsection{Data from 2013 and 2014}
\label{subsec:2013/4}
The quasar J1237+0647 was observed in visitor mode in May 2013, program 091.A-0124(A), and in service mode in the period March-June 2014, program 093.A-0373(A). The total integration time amounted to \mbox{11.5 hrs}. Each exposure was nominally 4800 s in duration and was expected to deliver a S/N of \mbox{13 per 2.5\kms per pixel} at \mbox{400 nm} and a resolving power \emph{R} of \mbox{$\sim$40~000}, with a seeing of \mbox{0.8 arcsec}, a slit of \mbox{1.0 arcsec}, an airmass of 1.2 and \mbox{2$\times$2} binning. Due to bad weather conditions, four exposures were shorter than scheduled and, therefore, the predicted S/N level could not be reached in those cases. 

Each of the quasar exposures had an `attached' standard ThAr calibration, which is a lamp exposure taken immediately after the science exposure leaving all the instrument parameters unchanged, and was immediately followed by a `supercalibration' exposure without allowing any grating reset. The supercalibration is a method to quantify any long-range wavelength distortion and it involves observing objects with a solar-like spectrum, like asteroids or `solar twin' stars (see Section \ref{sec:supercali}).

\subsection{Data from 2009}
\label{subsec:2009}
Part of the total dataset used in this work was retrieved from the ESO data archive\footnote{\url{http://archive.eso.org/eso/eso_archive_main.html}}. This was the case for observations from programs 082.A-0544(A) and 083.A-0454(A), which were carried out in March-April 2009 and were reported by \cite{Noterdaeme2010}. The main difference with the most recent observations of J1237+0647 is that these exposures only had the regular `non-attached' ThAr calibrations taken at the end of each night and dedicated observations of supercalibration targets were not carried out. These observations of \mbox{J1237+0647} from 2009 added \mbox{8.5 hrs} of integration on the target.

\section{Method}
\label{sec:method}
\subsection{Theory}
\label{subsec:theory}
\cite{Thompson1975} originally proposed that a possible temporal variation of the proton-to-electron mass ratio \mbox{$\mu = m_{p}/m_{e}$} can be detected using absorption spectra of molecular hydrogen observed at high redshift. The observed wavelength $\lambda^{z}_{i}$ of the $\emph{i}$-th transition will show a shift given by:
\begin{equation}
\lambda^{z}_{i}=\lambda^{0}_{i}(1+\zab)(1+K_{i}\frac{\Delta\mu}{\mu}),
\label{wl_shift}
\end{equation}
where $\lambda^{0}_{i}$ is the rest wavelength, $\zab$ is the redshift where the absorption occurs, $\Delta\mu \equiv \mu_{z} - \mu_{0}$ is the difference between the proton-to-electron mass ratio in the system observed and the value measured in the laboratory, and $K_{i}$ is the sensitivity coefficient which, for a given transition, determines the shifting power and sign due to varying $\mu$. These coefficients are defined as
\begin{equation}
K_{i} = \frac{\mathrm d \ln\lambda_{i}}{\mathrm d \ln\mu}.
\label{kcoeff}
\end{equation}
The sensitivity coefficients for the H$_{2}$ molecule used in this analysis were calculated within a semi-empirical framework by \cite{Ubachs2007}, including effects beyond the Born-Oppenheimer approximation.

The H$_{2}$ laboratory wavelengths, $\lambda^{0}_{i}$, used as a reference in this analysis were measured by \cite{Salumbides2008} with a fractional wavelength accuracy $\Delta\lambda/\lambda\sim5\times10^{-8}$, and they can be considered exact in comparison with the uncertainties of the lines in the quasar spectrum. The complete list of laboratory data required for fitting the H$_{2}$ absorption lines observed in this work, including rest wavelengths $\lambda^{0}_{i}$, oscillator strengths \emph{f}$_{i}$, damping coefficients $\Gamma_{i}$ and the sensitivity coefficients \emph{K}$_{i}$, was compiled by \cite{Malec2010}. 

\subsection{Fitting Method}
\label{subsec:fitmethod}
A comprehensive fitting method \mbox{\citep{Malec2010,King2011,Bagdonaite2014}}, which involves a simultaneous treatment of all the lines, was used to model the H$_{2}$ spectrum. Furthermore this technique allows to tie part of the fitting parameters together, resulting in a smaller number of free parameters. Simultaneous fitting and parameter tying allows to include also those H$_{2}$ transitions that are overlapped by metal or \mbox{H \textrm{\sc{i}} }lines from the Lyman $\alpha$ forest which can be modelled at the same time.

The program used is the non-linear least-squares Voigt profile fitting program \textsc{vpfit}\footnote{\url{http://www.ast.cam.ac.uk/~rfc/vpfit.html}}, developed specifically for quasar spectra analysis. A Voigt profile is obtained  from the convolution of a Lorentzian profile, which describes the natural line broadening of the transitions and is specific for each molecular transition considered, and a Gaussian profile, which reflects the physical conditions within the absorbing clouds by describing the velocities, both thermal and turbulent, of the molecules. These two profiles are eventually convolved with a model for the instrumental profile, which is assumed to be Gaussian.

In \textsc{vpfit}, the profile of each velocity component, of each transition, is described by three free parameters: the column density \emph{N}, the redshift \emph{z} at which the absorption occurs, and the Doppler width \emph{b}. These parameters are used in addition to the molecular physics input for every \emph{i}-th transition, like $\lambda^{0}_{i}$, \emph{f}$_{i}$, whose product with \emph{N} gives the optical depth of the absorption line, and $\Gamma_{i}$, which defines the lifetime broadening. Transitions probing the same rotational ground state share the same population, and as a consequence they are all described using the column density \emph{N}$_{J}$. An important underlying assumption is that absorbing features at the same redshift originate from the same cloud, hence they share certain parameters related to the physical conditions of the cloud. In particular, they are assumed to share the redshift \emph{z} and the width \emph{b}, which reflects the turbulent motion and the kinetic temperature of the gas in the cloud. This is achieved by tying in \textsc{vpfit} the \emph{z} and \emph{b} parameters of all the absorption lines associated with the gas in a single absorbing cloud.

The fitting program finds the best match between the data and the model iteratively by minimizing the $\chi^{2}$ parameter. The initial values of the free parameters are inserted by the user. Then, in each iteration, \textsc{vpfit}  changes the values of the free parameters of the model and checks the changes in relative $\chi^{2}$, reporting convergence once a stopping criterion, which is user-defined, is met.

Since introducing too many free parameters may lead to an overfit of the spectrum, models with a different number of velocity components per H$_{2}$ transition were compared using the $\chi^{2}_{\nu}$ and the Akaike Information Criterion~\citep[\emph{AICC},][]{Akaike1974} parameter. In particular, the latter parameter is defined as
\begin{equation}
AICC = \chi^{2} + 2p + \frac{2p(p+1)}{n-p-1}
\end{equation}
where \emph{p} is the number of free parameters and \emph{n} is the number of spectral points included in the fit. $\Delta$\emph{AICC}$>5$ and $\Delta$\emph{AICC}$>10$ are considered to be a strong and a very strong evidence respectively that the model with the lower \emph{AICC} is statistically preferred.

Multiple fits, with a different number of velocity components each, were performed in order to develop a robust absorption model. Only after the results of the fitting process are stable, the possible variation of the proton-to-electron mass ratio expressed by the variable $\Delta\mu/\mu$ is invoked as an extra free parameter in the fit besides \emph{N$_{J}$}, \emph{z} and \emph{b}. This fourth parameter is not introduced earlier during the fitting process to avoid that any over/underfitting can be compensated by an artificial variation of $\mu$. The program derives the value of $\Delta\mu/\mu$ according to Eq. (\ref{wl_shift}), therewith also using the sensitivity coefficients of Eq. (\ref{kcoeff}). The consistency of the derived $\Delta\mu/\mu$ constraint is then tested for sensitivity to a number of assumptions made in the fitting process. The results will be discussed in detail in Section \ref{subsec:constraint}.

\section{Analysis}
\label{sec:analysis}
Quasar J1237+0647 is located at redshift \emph{z}=2.78, thereby defining the extent of the \mbox{Lyman $\alpha$} forest, which comprises multiple series of neutral hydrogen transitions arising from the intergalactic medium at all redshifts below the quasar redshift. Toward the quasar there is one major absorption system at redshift $z=2.69$, which is a sub-damped \mbox{Lyman $\alpha$} system (DLA) with a neutral hydrogen column density of \mbox{$\log N=20.0\pm0.15$ cm$^{-2}$}, that features many absorption lines from atomic, ionized and molecular species \mbox{\citep{Noterdaeme2010}}.

\begin{figure}
\centering
\includegraphics[width=0.5\textwidth]{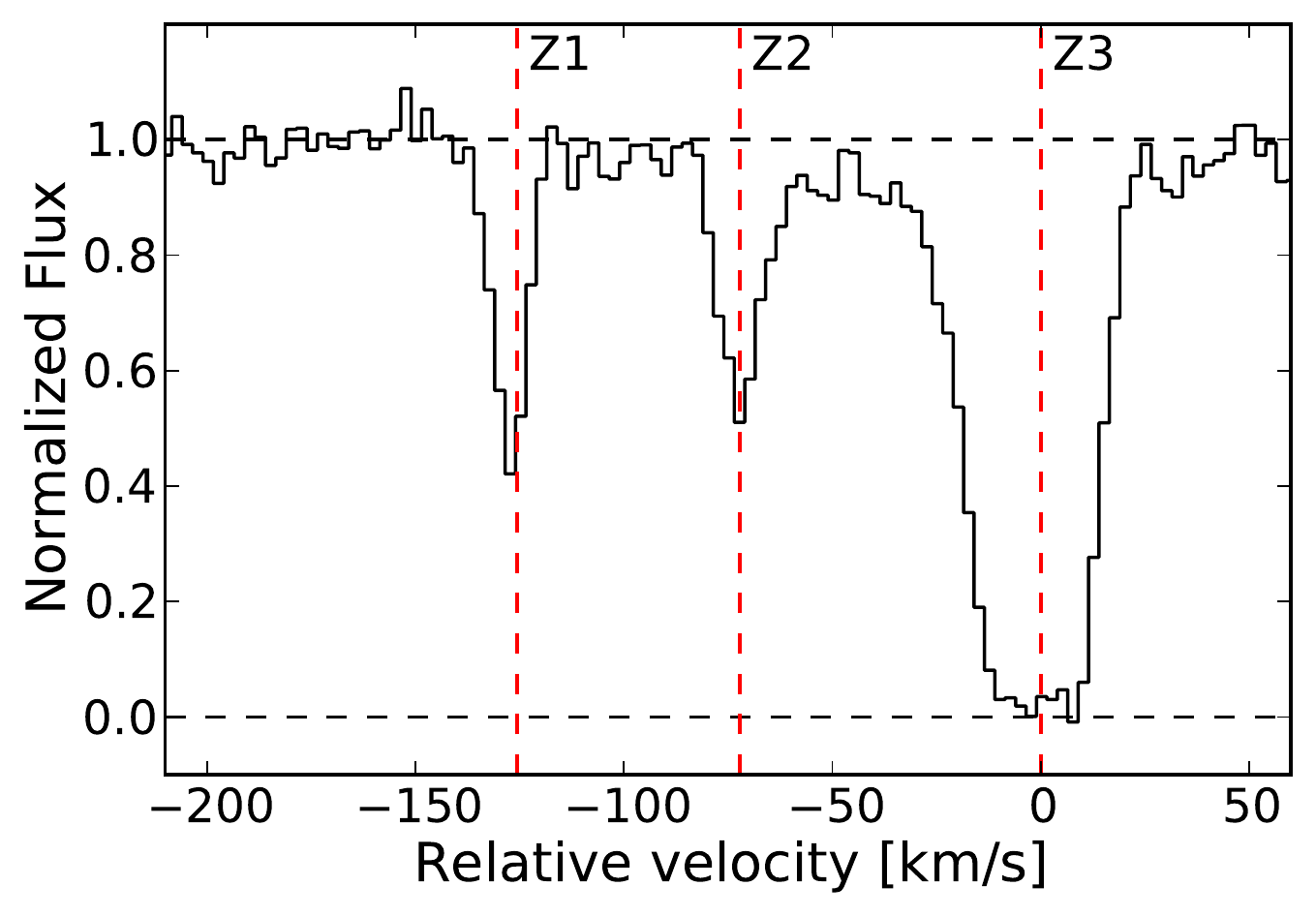}
\caption{Typical line profile for the three absorption features toward J1237+0647. The transition plotted is L1P(3) at \mbox{$\lambda_{0}=109.978$ nm}. The vertical dashed lines (red) indicate the positions of the three H$_{2}$ spectral features fitted in this study. The velocity frame is centered on the strongest feature Z3.}
\label{lprof}
\end{figure}

Molecular hydrogen absorption lines associated with the DLA are spread in the range from 352.1 to \mbox{409.5 nm}. H$_{2}$ is found in three absorption features at redshift \mbox{\emph{z}=2.688001}, \mbox{\emph{z}=2.68868} and \mbox{\emph{z}=2.68955}; their velocities relative to the redshift of the strongest feature at \mbox{\emph{z}=2.68955} are respectively \mbox{\emph{v}=$-125$\kms}, hereafter Z1, \mbox{\emph{v}=$-72$\kms}, hereafter Z2, and \mbox{\emph{v}=0\kms}, hereafter Z3. The presence of these three features close to each other results in the line profiles shown in Fig. \ref{lprof}. The presence of clearly separated velocity features is not uncommon and has been observed in various systems in the past. The number of such components can vary from one up to the seven distinct features observed in the absorbing system toward quasar Q2348--011 \mbox{\citep{Petitjean2006,Bagdonaite2012}}.

Z3 has the highest column density for every \emph{J}-level, resulting in heavily saturated absorption features for low \emph{J}-levels with \mbox{\emph{J}$\le1$}. Furthermore, Z3 shows also deuterated molecular hydrogen (HD) absorption, where only the \mbox{\emph{J}=0 level} was detected \mbox{\citep{Noterdaeme2010}}. The HD molecule is sensitive to a variation of $\mu$ \mbox{\citep{Ivanov2008,Ivanov2010}}, and its detected transitions were included in the present $\mu$-variation analysis.

\subsection{Creating an absorption model}
\label{subsec:model}

\subsubsection{Selecting and fitting spectral regions}
\label{ssubsec:regions}
Given the broad absorption profile, spanning \mbox{$\sim150$\kms}, overlaps between profiles of different molecular hydrogen transitions are common as well as overlaps with the H$\,$\textsc{i} transitions from the Lyman $\alpha$ forest. Within the comprehensive fitting method, all kinds of overlaps can be handled. However, no relevant information was gained in the case of a complete overlap with a saturated H$\,$\textsc{i} line, hence such absorption features were not included in this study. Less frequently, some H$_{2}$ features partially overlap with narrow metal transitions. However, these lines were not excluded from the dataset considered in this work. A large number of metal absorption features were identified by \mbox{\cite{Noterdaeme2010}}. 

All the 137 potentially useful H$_{2}$ and HD transitions were contained in 66 spectral regions, in the range \mbox{351.0--413.5 nm}, which were selected trying to avoid, whenever possible, \mbox{H$\,$\textsc{i}} transitions and too-heavily saturated low \mbox{\emph{J}-level} H$_{2}$ absorption features in Z3. The useful H$_{2}$ transitions, probing six different rotational quantum states with \emph{J}=0--5, are listed in Table \ref{transitions}. Absorption features originating at the same redshift had both the \emph{z} and the \emph{b} parameters tied together, while the column densities \emph{N}$_{J}$ were allowed to vary independently from each other. Z3 shows saturated H$_{2}$ lines, particularly in the low \emph{J}-levels, but there is an apparent evidence of a partial coverage with a non-zero residual flux at the base of its absorption features. To take into account this effect, a zero-level correction was included as a free parameter in the fit for each H$_{2}$ region.
\begin{table*}
\centering
\caption{List of the molecular transitions used in this analysis.}
\label{transitions}
\begin{tabular}{cp{7.5cm}p{5cm}c}
\hline
\emph{J}-level & Lyman transitions & Werner transitions & n$_{\textrm{transitions}}$ \\
\hline
\hline
H$_{2}$ \emph{J} = 0 & L0R(0), L2R(0), L3R(0), L4R(0), L6R(0), L7R(0), L8R(0), L10R(0) & W0R(0), W1R(0) & 10 \\
H$_{2}$ \emph{J} = 1 & L0P(1), L0R(1), L1P(1), L1R(1), L2P(1), L2R(1), L3P(1), L7P(1), L9R(1), L10R(1), L12R(1) & W0R(1), W1R(1) & 13 \\
H$_{2}$ \emph{J} = 2 & L0P(2), L0R(2), L1P(2), L1R(2), L2P(2), L2R(2), L3P(2), L4P(2), L4R(2), L5P(2), L6P(2), L7P(2), L7R(2), L8P(2), L9P(2), L10P(2), L10R(2), L11P(2), L11R(2), L13P(2), L13R(2) & W0P(2), W0Q(2), W1Q(2), W2P(2), W2Q(2), W2R(2) & 27 \\
H$_{2}$ \emph{J} = 3 & L0P(3), L0R(3), L1P(3), L1R(3), L2R(3), L3P(3), L3R(3), L5P(3), L5R(3), L6P(3), L6R(3), L7P(3), L7R(3), L8P(3), L8R(3), L9R(3), L10P(3), L12R(3), L13P(3), L13R(3), L14P(3) & W0P(3), W0Q(3), W1R(3), W2P(3), W2Q(3), W2R(3), W3P(3) & 28\\
H$_{2}$ \emph{J} = 4 & L0P(4), L0R(4), L1P(4), L1R(4), L2R(4), L3P(4), L3R(4), L4P(4), L4R(4), L5R(4), L6P(4), L7P(4), L7R(4), L8P(4), L9P(4), L9R(4), L11P(4), L11R(4), L12R(4), L13P(4), L14P(4) & W0P(4), W0Q(4), W0R(4), W1P(4), W2Q(4), W2R(4), W3Q(4) &  28\\
H$_{2}$ \emph{J} = 5 & L0R(5), L1P(5), L2P(5), L2R(5), L3P(5), L3R(5), L4P(5), L5P(5), L5R(5), L7P(5), L8P(5), L9P(5), L10P(5), L11R(5), L12R(5), L13P(5), L14P(5), L15P(5) & W0P(5), W0R(5), W1Q(5), W2Q(5), W2R(5), W3P(5) & 24 \\
\hline
HD \emph{J} = 0 & L3R0, L5R0, L8R0, L12R0, L15R0 & W1R0, W3R0 & 7 \\
\hline
\multicolumn{3}{l}{Total number of lines} & 137 \\
\hline
\end{tabular}
\end{table*}

In order to build a robust model, the neutral hydrogen absorption lines occurring in the Lyman $\alpha$ forest and included in the selected regions must be accounted for. The \mbox{H$\,$\textsc{i}} transitions were modelled by assigning to each of them a set of free parameters (\emph{N}, \emph{z}, \emph{b}) in \textsc{vpfit}. Initial guesses of the free parameters were user-provided for each H$_{2}$, HD and \mbox{H$\,$\textsc{i}} transition.

\subsubsection{Extra velocity components}
\label{ssubsec:extravcs}
The H$_{2}$ absorption spectrum toward J1237+0647 displays three clearly distinct features in velocity space. In order to fit the spectrum as accurately as possible, it is imperative to consider possible underlying substructures  in velocity space. Physically this may relate to the existence of closely separated distinct clouds, or inhomogeneities in the clouds in the line-of-sight. To investigate such underlying velocity structure, the fitting residuals from the 66 selected regions were normalized, shifted to a common velocity scale and averaged together, creating a composite residual spectrum \mbox{\citep[CRS,][]{Malec2010}}. Only the H$_{2}$ absorption features that were not overlapped by \mbox{H$\,$\textsc{i}} features were included in the CRS, in order to avoid that the residuals of the \mbox{H$\,$\textsc{i}} model could introduce any spurious effect. The CRS, by combining the residuals of many transitions, highlighted any over/underfitted structure in the absorption features.

\begin{figure}
\includegraphics[width=\columnwidth]{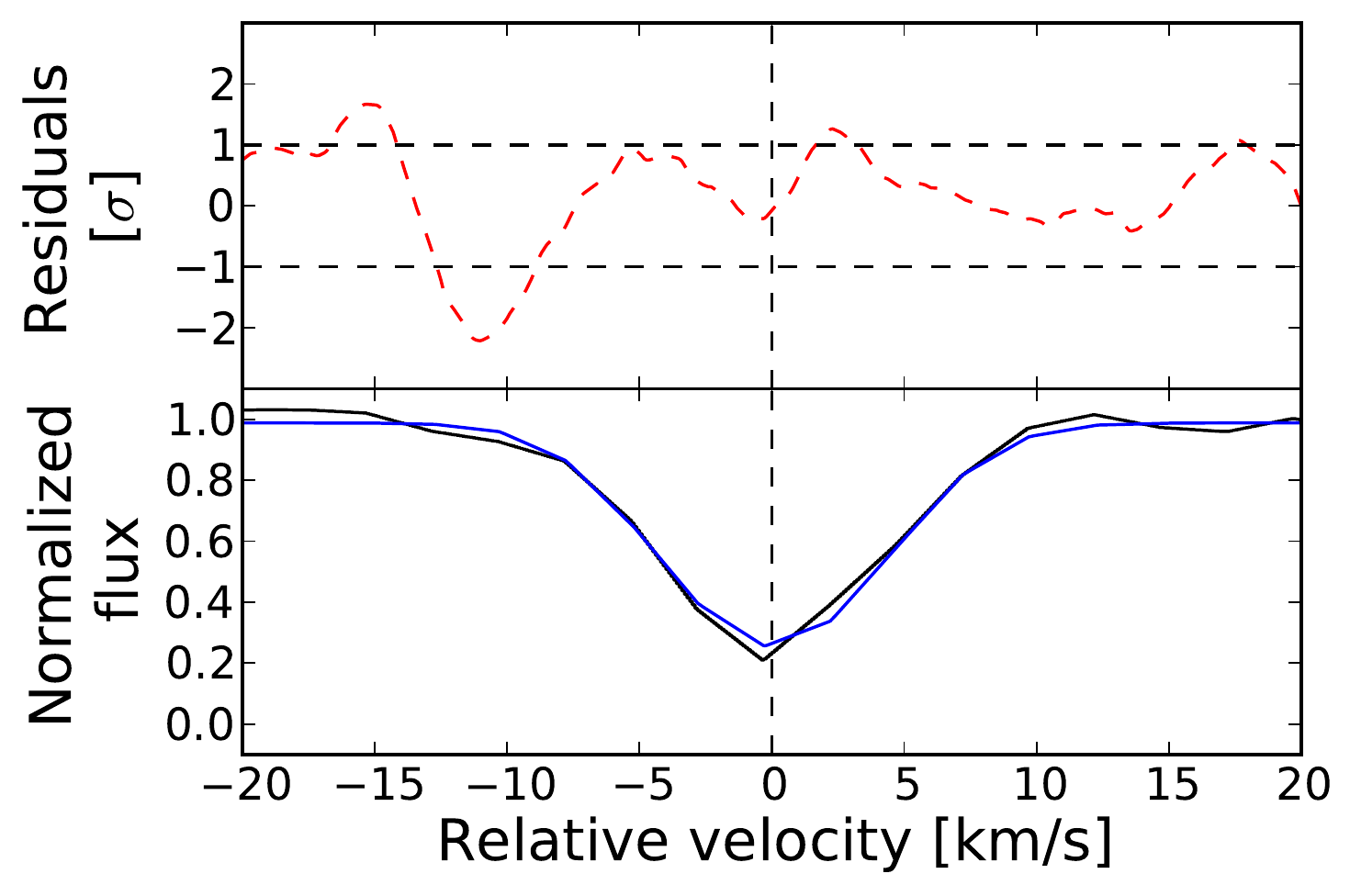}
\caption{Top panel: normalized composite residual spectrum for Z1 formed from 30 H$_{2}$ absorption features. The dashed lines represent the $\pm 1\sigma$ boundaries. Bottom panel: an example transition shown on the same velocity scale. The velocity scale is centred at \mbox{$z = 2.688009$}. The vertical dashed line shows the position of the velocity component.}
\label{res1}
\end{figure}

\begin{figure}
\includegraphics[width=\columnwidth]{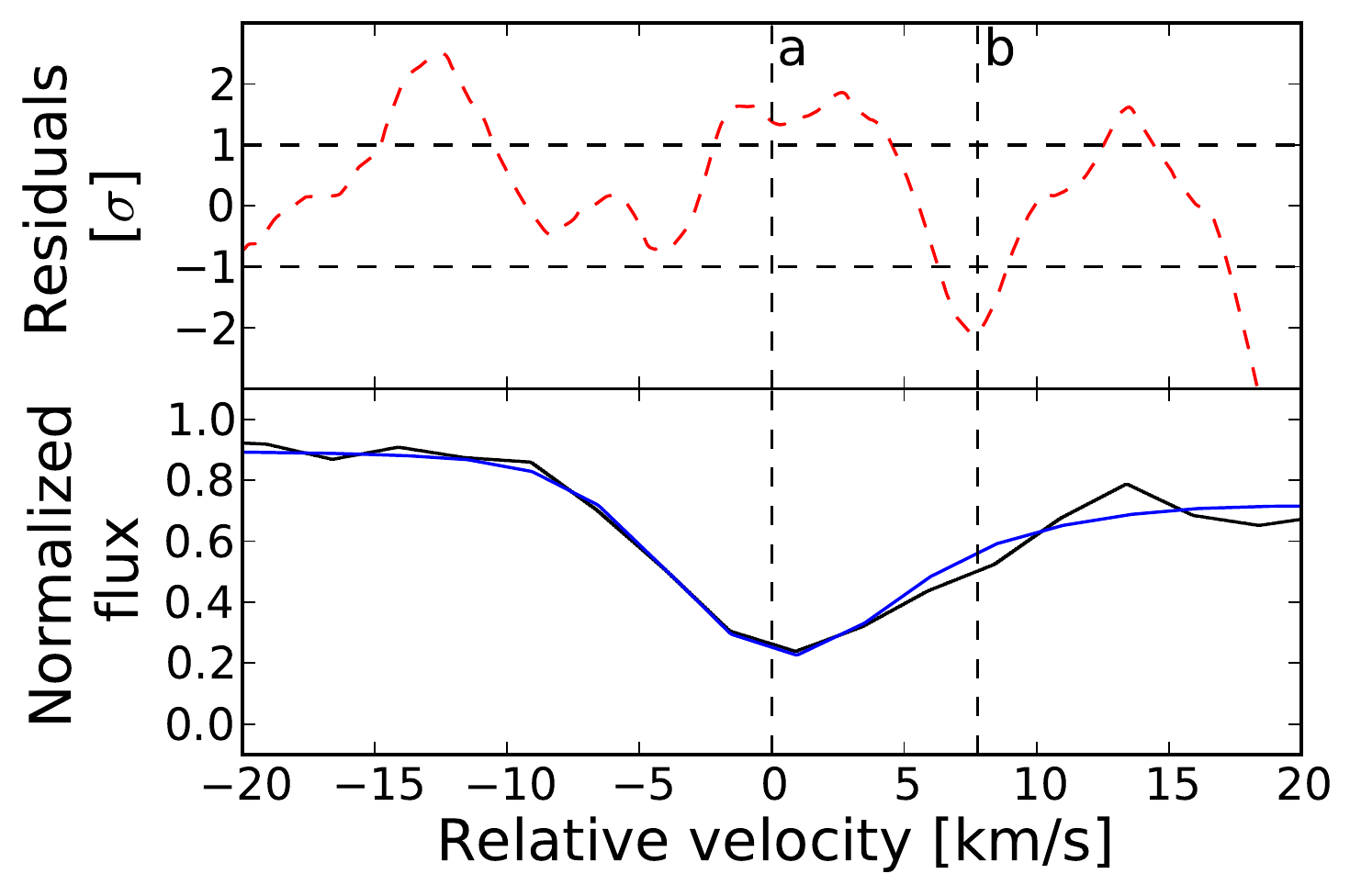}
\caption{Top panel: normalized composite residual spectrum for Z2 formed from 33 H$_{2}$ absorption features obtained from the 2 VCs model. The dashed lines represent the $\pm 1\sigma$ boundaries. Bottom panel: an example transition shown on the same velocity scale. The velocity scale is centred at \mbox{$z = 2.688664$}, on the strongest velocity component. The vertical dashed lines show the positions of the two velocity components at $v = 0$\kms (a) and +7\kms (b).}
\label{res2}
\end{figure}

\begin{figure}
\includegraphics[width=\columnwidth]{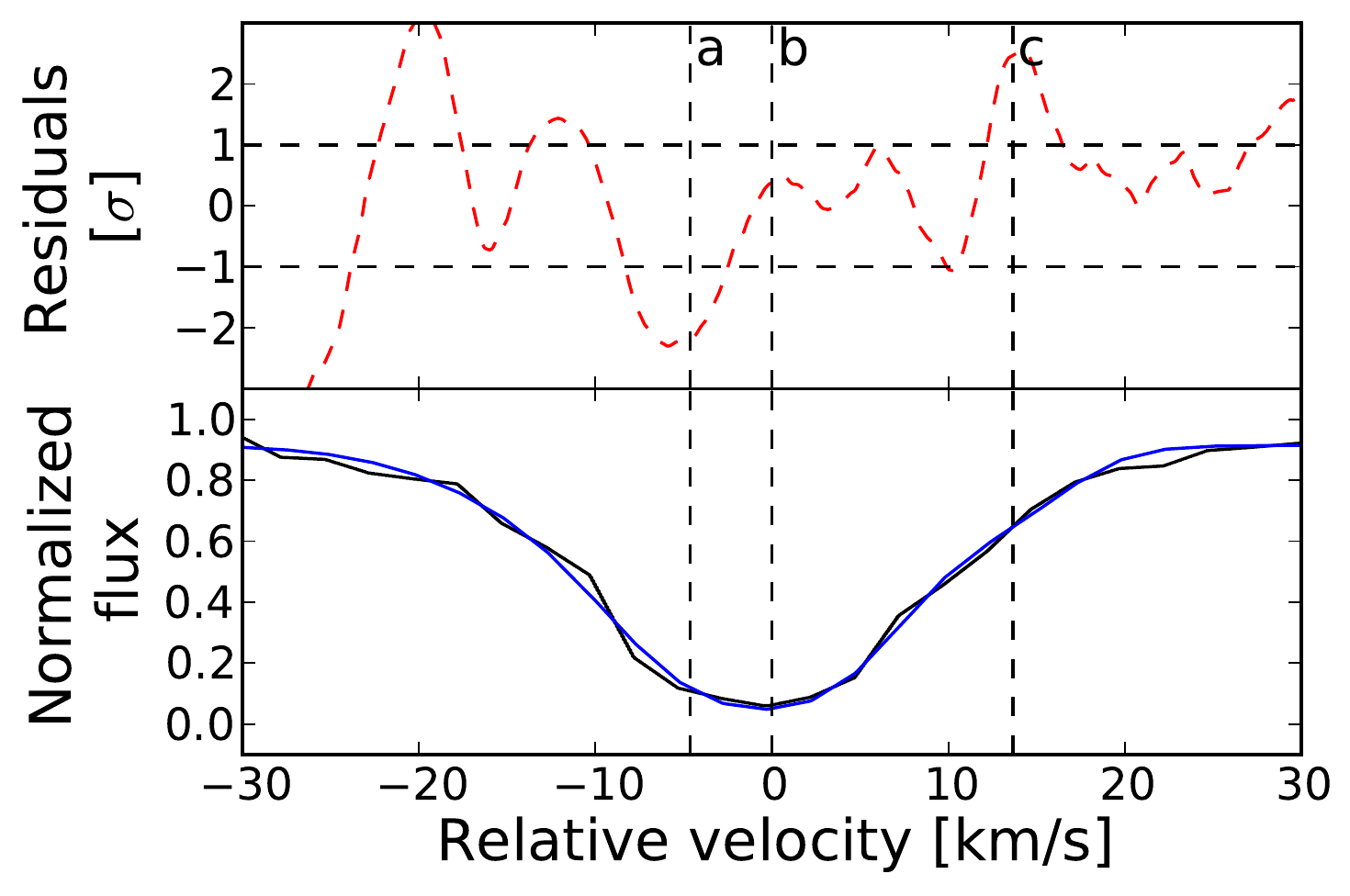}
\caption{Top panel: normalized composite residual spectrum for Z3 formed from 33 H$_{2}$ absorption features obtained from the 3 VCs model. The dashed lines represent the $\pm 1\sigma$ boundaries. Bottom panel: an example transition shown on the same velocity scale. The velocity scale is centred at \mbox{$z=2.689555$}, on the strongest velocity component. The vertical dashed lines show the positions of the three velocity components at $v = -4$\kms (a), 0\kms (b) and +13\kms (c).}
\label{res3}
\end{figure}

\begin{figure*}
\centering
\includegraphics[width=2\columnwidth]{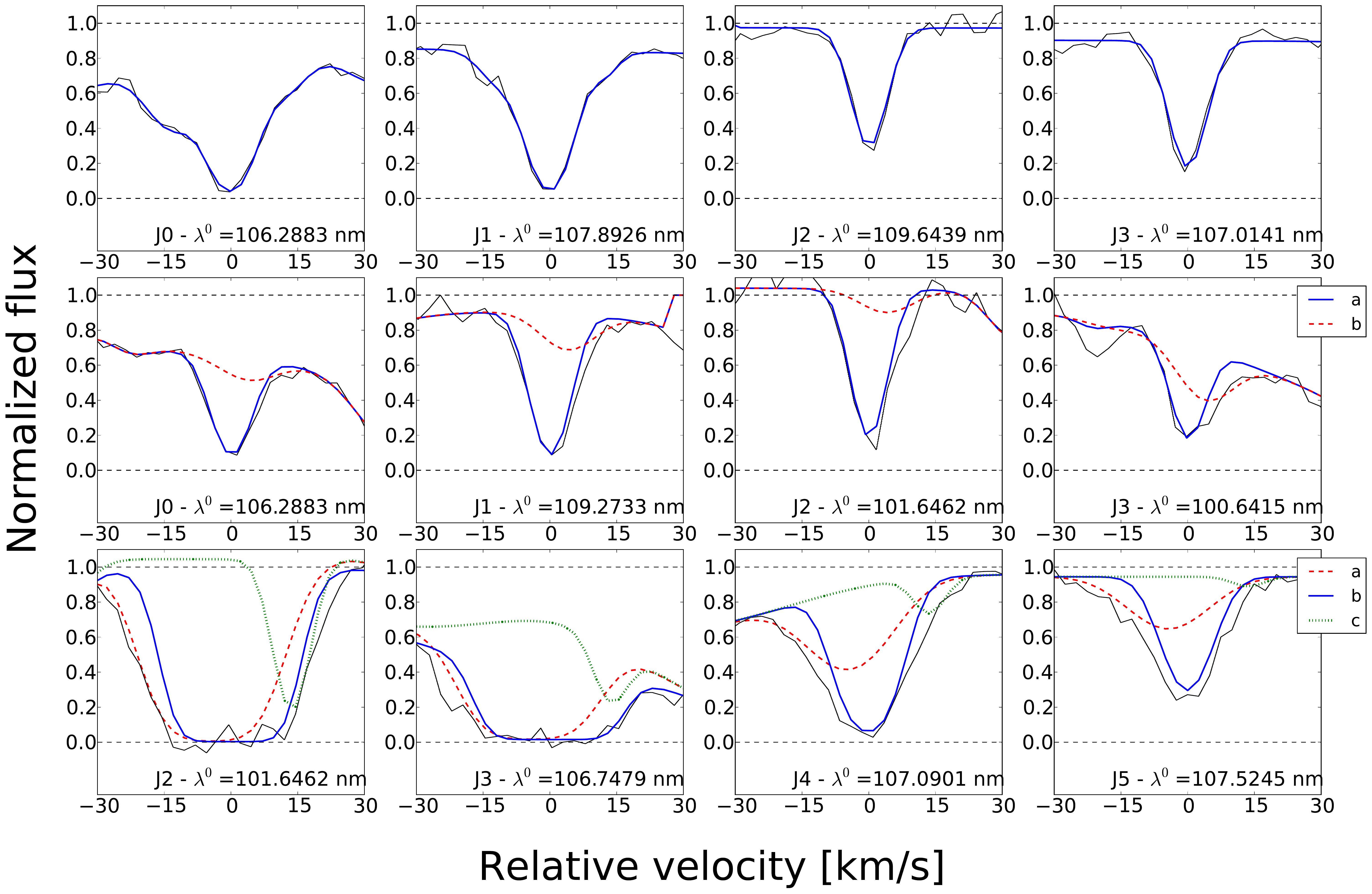}
\caption{Example transitions for every \emph{J}-level. Each velocity component is plotted against the fitted continuum. \mbox{H$\,$\textsc{i}} contributions to the line shapes are not highlighted. Top panel: \emph{J}-levels 0-4 for Z1 (blue), the velocity scale is centred at \emph{z}=2.688009. Middle panel: \emph{J}-levels 0-4 for Z2. The velocity component at \emph{z}=2.688664, Z2a, is depicted with the (blue) solid line and the one at \emph{z}=2.688717, Z2b, with the (red) dashed line. The velocity scale is centred on Z2a. Bottom panel: \emph{J}-levels 2-5 for Z3. The three velocity components at redshifts \emph{z}=2.689498, Z3a, 2.689555, Z3b, and 2.689723, Z3c, are shown with the dashed (red), solid (blue) and dotted (green) line respectively. The velocity scale is centred on Z3b.}
\label{abs1}
\end{figure*}

Z1 could be described using only one velocity component (VC), while indications of underfitting were observed in the CRS for absorption features Z2 and Z3, hence they were modelled by adding extra VCs. They were added to the model until \textsc{vpfit} started to reject them. Then the models with a different number of VCs  were compared using the $\chi^{2}$ and the \emph{AICC} parameters, in order to decide which one better fitted the data. The CRS for the three absorption features Z1, Z2 and Z3 are shown in Figs. \ref{res1}--\ref{res3}. Z2 could be modelled using two VCs, labelled as `a' and `b' in Fig. \ref{res2}, and Z3 was modelled using three VCs, labelled as `a', `b' and `c' in Fig. \ref{res3}. Parameters for the model are summarized in Table \ref{param} and the VCs are shown in Fig. \ref{abs1}. Previous studies found no strong evidences for a systematic shift in $\Delta\mu/\mu$ due to a segregation in different rotational levels \mbox{\citep{Malec2010,King2011}}, hence it was assumed that the velocity structure is the same among all the \emph{J}-levels in each absorption feature. The complete absorption model is shown in Section \ref{app:model}.

•
•
•

\begin{table*}
\centering
\caption{Column densities \emph{N$_{J}$}, redshift \emph{z} and Doppler parameters \emph{b} of the H$_{2}$ transitions in the three clearly distinguishable absorption features and in their underlying velocity components a, b, c. Levels with \emph{J} = 4 and 5 are too weak to be detected in Z1 and Z2. Levels with \emph{J} = 0 and 1 are heavily saturated in Z3 and they have been discarded from the model.}
\label{param}
\begin{tabular}{c|c|cc|ccc}
 & Z1 & \multicolumn{2}{c}{Z2} & \multicolumn{3}{c}{Z3} \\
 \hline
\emph{J}-levels & $\log{N_{J}}$ [cm$^{-2}$] & \multicolumn{2}{c}{$\log{N_{J}}$ [cm$^{-2}$]} & \multicolumn{3}{c}{$\log{N}_{J}$ [cm$^{-2}$]} \\
 & & a & b & a & b & c \\
\hline
\emph{J} = 0 & $15.74 \pm 0.11$ & $15.44 \pm 0.11$ & $13.78 \pm 0.17$ & \multicolumn{3}{c}{Discarded} \\
\emph{J} = 1 & $16.29 \pm 0.08$ & $16.23 \pm 0.11$ & $14.56 \pm 0.10$ & \multicolumn{3}{c}{Discarded} \\
\emph{J} = 2 & $15.35 \pm 0.04$ & $15.27 \pm 0.07$ & $13.90 \pm 0.15$ & $15.67 \pm 0.10$ & $17.64 \pm 0.10$ & $15.26 \pm 0.16$ \\
\emph{J} = 3 & $15.21 \pm 0.02$ & $14.88 \pm 0.05$ & $14.31 \pm 0.08$ & $15.66 \pm 0.11$ & $18.02 \pm0.06$ & $14.64 \pm 0.28$ \\
\emph{J} = 4 & \multicolumn{3}{c}{N/D} & $14.79 \pm 0.07$ & $15.34 \pm 0.04$ & $13.93 \pm 0.07$ \\
\emph{J} = 5 & \multicolumn{3}{c}{N/D} & $14.50 \pm 0.07$ & $14.79 \pm 0.04$ & $13.33 \pm 0.17$ \\
HD \emph{J} = 0 & \multicolumn{3}{c}{N/D} & N/D & $14.36 \pm 0.03$ & N/D \\
\hline
\emph{z} & 2.68801(06) & 2.68866(09) & 2.68872(82) & 2.68950(96) & 2.68955(14) & 2.68972(52) \\
\hline
\emph{b} [kms$^{-1}$] & $2.94 \pm 0.08$ & $2.32 \pm 0.13$ & $6.99 \pm 0.65$ & $11.36 \pm 0.48$ & $5.78 \pm 0.11$ & $2.54 \pm 0.48$ \\
\hline
\end{tabular}
\end{table*}

\subsection{Constraining $\Delta\mu/\mu$}
\label{subsec:constraint}
\begin{figure}
\centering
\includegraphics[width=\columnwidth]{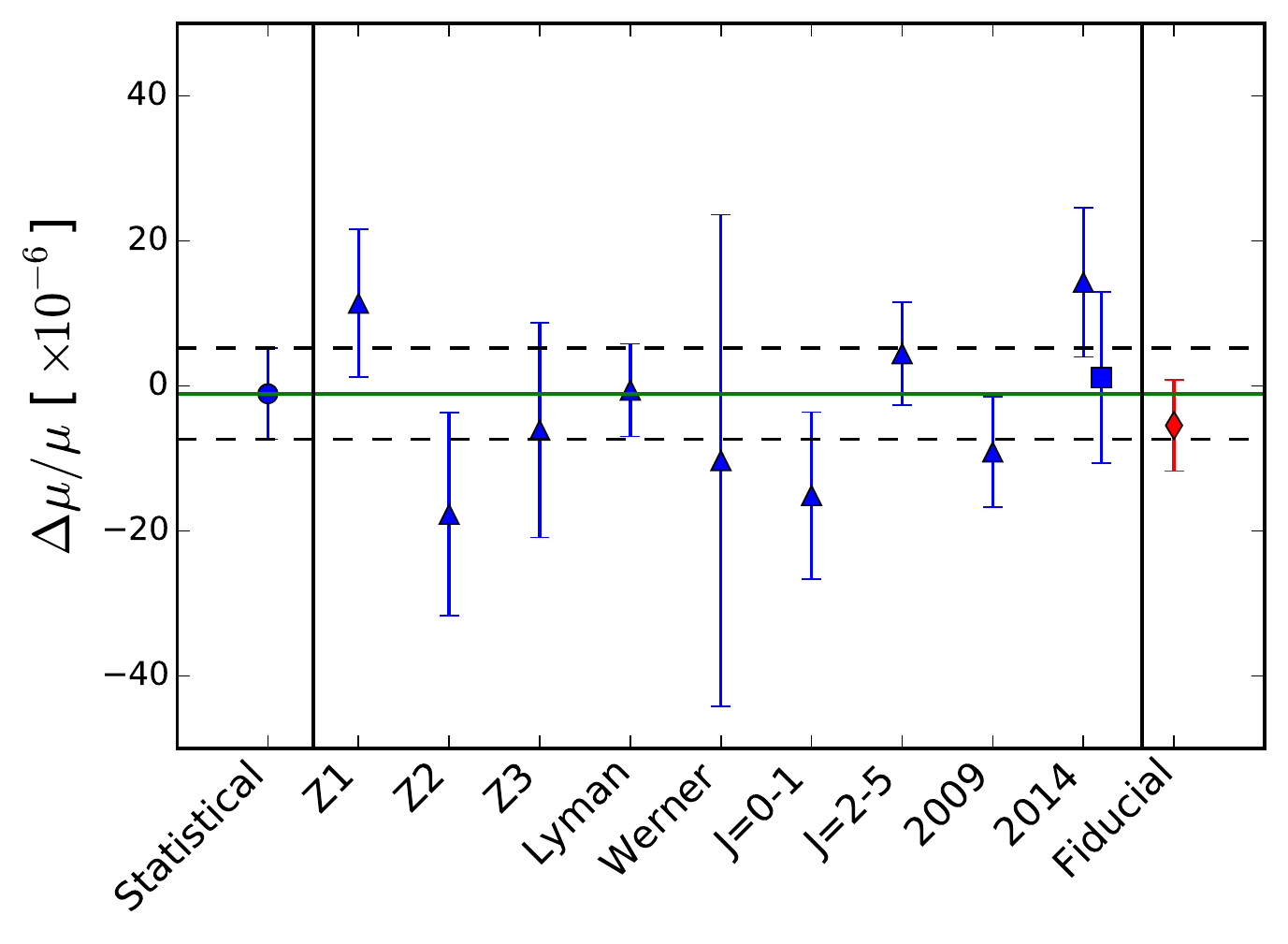}
\caption{$\Delta\mu/\mu$ constraints obtained from various tests (triangles) compared with the statistical value from the entire dataset (circle). The solid (green) horizontal line indicates the value of the statistical constraint and the two dashed lines represent its $1 \sigma$ statistical uncertainties. The square represents the $\Delta\mu/\mu$ value obtained only from the exposures taken in 2014 after applying to each of them a long-range distortion correction, as discussed in Section \ref{sec:supercali}. The diamond (red) represents the fiducial value obtained for the entire dataset (2009+2013+2014) after applying the distortion corrections.}
\label{tests}
\end{figure}

Once an optimal model was found, an extra free parameter, corresponding to $\Delta\mu/\mu$, was added to the H$_{2}$ transitions in \textsc{vpfit}. All the transitions were included in the fit, without any distinction between their bands, their \emph{J}-levels or absorbing clouds where they arose. The model returns \mbox{$\Delta\mu/\mu=(-1.1 \pm 6.3_{\textrm{\small{stat}}})\times10^{-6}$}. The statistical error reported here was derived from the appropriate diagonal term of the final covariance matrix for the fit. Given the absorption model, it represents only the uncertainty in $\Delta\mu/\mu$ stemming from the photon statistics of the quasar spectrum, i.e. its signal-to-noise ratio. However, to evaluate the robustness of the result, hereafter the statistical value of $\Delta\mu/\mu$, a variety of consistency tests and sources of systematic error which contribute to the overall uncertainty budget were explored, results of which are shown in Fig. \ref{tests} and discussed below.

\subsubsection{Isolating single absorbing features}
\label{ssubsec:singleabs}
Each of the absorbing features Z1, Z2 and Z3 can be investigated separately. The constraints on $\Delta\mu/\mu$ returned are \mbox{$\Delta\mu/\mu\vert_{\textrm{\small{Z1}}}=(11.4 \pm 10.2_{\textrm{\small{stat}}})\times10^{-6}$} for Z1, \mbox{$\Delta\mu/\mu\vert_{\textrm{\small{Z2}}}=(-17.7 \pm 14.0_{\textrm{\small{stat}}})\times10^{-6}$} for Z2 and \mbox{$\Delta\mu/\mu\vert_{\textrm{\small{Z3}}}=(-6.1 \pm 14.8_{\textrm{\small{stat}}})\times10^{-6}$} for Z3. The resulting constraints are in agreement with each other and with the statistical value for the entire dataset within $1 \sigma$ boundaries.

\subsubsection{Isolating Lyman and Werner transitions}
\label{ssubsec:lymanwerner}
All the H$_{2}$ transitions considered in this analysis belong either to the Lyman- or to the Werner band systems. Any long-range distortion of the wavelength scale will be nearly degenerate with a shift of the molecular hydrogen lines due to a non-zero value of $\Delta\mu/\mu$. This degeneracy is partly broken by fitting together H$_{2}$ transitions from both Lyman and Werner bands. The effect of such a distortion can be investigated by fitting separately transitions from the two band systems.

The values returned in this way are \mbox{$\Delta\mu/\mu\vert_{\textrm{\small{L}}} = (-0.6 \pm 6.4_{\textrm{\small{\small{stat}}}}) \times 10^{-6}$} for Lyman transitions and \mbox{$\Delta\mu/\mu\vert_{\textrm{\small{W}}} = (-10.3 \pm 33.9_{\textrm{\small{stat}}}) \times 10^{-6}$} for Werner transitions. The larger error of the value returned by Werner transitions is due to three effects. Firstly, there are fewer Werner transitions than Lyman transitions in the spectrum, with a ratio between the former and the latter of $\sim$1:4. Then the spread of the sensitivity coefficients is larger for Lyman transitions, \mbox{$\Delta K_{\textrm{\small{L}}} = 0.092$}, than the one of Werner transitions, \mbox{$\Delta K_{\textrm{\small{W}}} = 0.054$}. Finally, the Werner transitions fall in the bluest part of the spectrum, where the S/N is much lower than in the region where Lyman transitions are. Such differences lead the statistical value of $\Delta\mu/\mu$ to be dominated by the Lyman transitions.

\subsubsection{Isolating cold and warm states}
\label{ssubsec:coldwarm}
Using \textsc{vpfit} it is possible to calculate a different value of $\Delta\mu/\mu$ for transitions originating in each \emph{J}-level, in order to obtain the relative contributions of different levels to the statistical value. This is particularly useful to investigate the impact of possible temperature inhomogeneities in the absorbing clouds. Due to the para-ortho distribution, the \emph{J}$=1$ rotational state is significantly populated even at the lowest temperatures \mbox{\citep{Ubachs2007}}. As a consequence, the H$_{2}$ transitions can be divided into two sets: rotational states with \emph{J}=0,1 (cold) and with \emph{J}$\ge2$ (warm).

The constraints returned by the model are \mbox{$\Delta\mu/\mu\vert_{\textrm{\small{cold}}} = (-15.1 \pm 11.5_{\textrm{\small{stat}}}) \times 10^{-6}$} for the cold states and \mbox{$\Delta\mu/\mu\vert_{\textrm{\small{warm}}} = (4.4 \pm 7.1_{\textrm{\small{stat}}}) \times 10^{-6}$} for the warm states. The two values match within their uncertainties, which means that any temperature inhomogeneity in the absorbing clouds does not have a significant impact on the $\mu$-variation.

Furthermore, the slightly larger error derived from the cold states reflects the fact that levels with \emph{J}$\le3$ are saturated in Z3 and, in particular, cold states are heavily saturated. As a consequence of the saturation, there are fewer constraining pixels contributing to the fitting process and there are fewer transitions in the cold states dataset.

\subsubsection{Separating exposures from 2009}
\label{ssubsec:20092014}
The dataset used in this work can be divided between the exposures taken in 2013 and 2014, which had an attached ThAr calibration and a supercalibration, and the exposures taken in 2009, retrieved from the ESO archive, which only had regular non-attached ThAr calibrations. In order to see how the absence of the individual calibrations affects the statistical value, two values for $\Delta\mu/\mu$ were derived from the two subsets.

The constraints obtained in this way are \mbox{$\Delta\mu/\mu\vert_{\textrm{\small{2009}}} = (-9.1 \pm 7.6_{\textrm{\small{stat}}}) \times 10^{-6}$} for the exposures taken in 2009 and \mbox{$\Delta\mu/\mu\vert_{\textrm{\small{2014}}} = (14.3 \pm 10.3_{\textrm{\small{stat}}}) \times 10^{-6}$} for the exposures taken in 2013 and 2014. The two values do not agree within their combined 1$\sigma$ statistical uncertainties, although they do within \mbox{$1.5 \sigma$}.
However, this comparison does not take into account the long-range distortion effect, which will be discussed in the next section.

\section{supercalibrations}
\label{sec:supercali}
In order to constrain a variation of $\mu$, it is crucial to accurately wavelength calibrate the quasar observations. Attempts to perform advanced tests of the effective accuracy of UVES were performed by \mbox{\cite{Molaro2008}} by comparing a UVES-recorded asteroid spectrum with a highly accurate solar reference spectrum. These measurements revealed zero offsets up to \mbox{$\approx$ 50 ms$^{-1}$}, likely due to a non-uniform slit illumination, but found no evidence for a wavelength-dependent velocity shift in the spectrum. Similar studies, that are now referred to as `supercalibrations', showed that spectra taken with UVES and HIRES spectrographs suffer from intra-order calibration shifts on scales of single echelle orders but they did not uncover any evidence for a long-range distortion in the spectra on the scale of the spectrographs arms \mbox{\citep{Whitmore2010,Griest2010}}. \mbox{\cite{Rahmani2013}} detected for the first time a long-range distortion in some UVES spectra up \mbox{to $\sim$ 400 ms$^{-1}$} on a scale \mbox{of $\sim$ 600 \AA}. More recently, \mbox{\cite{Whitmore2015}} found that significant wavelength calibration distortions, both intra-order and long-range, are ubiquitous across UVES and HIRES history, most likely due to a different light path of the astronomical object and the ThAr lamp within the instrument.

The presence of long-range distortions between the attached ThAr calibration and the main target spectrum velocity scales introduces a wavelength-dependent shift that mimics a non-zero $\Delta\mu/\mu$. This happens because the sensitivity coefficients \emph{K$_{i}$} decrease with increasing wavelength (see Fig. 1 of \mbox{\cite{Bagdonaite2014}}) and a long-range distortion of the wavelength scale would produce a strong systematic effect which will be nearly degenerate with the effect produced by a variation of $\mu$. In principle, this degeneration can be broken by fitting Lyman and Werner transitions together; however, the paucity of Werner transitions and the low S/N in the spectral region where they fall prevent the degeneracy to be broken in this way, as illustrated by \mbox{\cite{Malec2010}} and discussed in Section \ref{ssubsec:lymanwerner}. 
Over most of UVES's history, the sign of measured long-range distortions was such that they spuriously pushed $\Delta\mu/\mu$ to more positive values \mbox{\citep{Whitmore2015}}.

\subsection{Method}
\label{subsec:supercalimethod}
The first supercalibration performed with UVES consisted of a comparison between asteroid spectra and a Fourier Transform Spectrometer (FTS) solar atlas \mbox{\citep{Molaro2008}}. \mbox{\cite{Rahmani2013}} cross-correlated asteroid spectra with a laboratory solar spectrum and, more recently, \mbox{\cite{Whitmore2015}} improved the technique by forward modelling the FTS spectra to match observed asteroid spectra, thereby allowing information to be derived from much shorter spectral ranges, and by showing that the same technique can be applied on solar twin stars as well. Their method is used in this study to supercalibrate the quasar spectrum.

The supercalibration process consists in the comparison of a ThAr calibrated spectrum with a reference spectrum from a FTS, whose frequency scale is expected to be much more accurate than the UVES one, hence it is considered `absolutely calibrated' for the purpose of this study. The reference used is the FTS vacuum solar spectrum from \cite{Chance2010}\footnote{Available at \url{http://kurucz.harvard.edu/sun/irradiance2005/irradthu.dat}}. The targets for the supercalibrations were asteroids, which reflect the solar light and hence have the same spectrum as the Sun, and `solar twin' stars, i.e. objects whose spectra are almost identical to the solar one \mbox{\citep{Melendez2009,Datson2014}}. Since the supercalibration targets are astronomical objects, no changes are required in the focus or in the slit alignment.

The two spectra were divided into small regions \mbox{of $\approx$ 500\kms} each, which corresponds to \mbox{$\approx$ 8 \AA}. Their small size allows for sampling each echelle order with \mbox{$\approx$ 10} regions, and for identifying any relative velocity distortion down to \mbox{$\sim$ 30 ms$^{-1}$} \mbox{\citep{Whitmore2015}}. The spectral regions of the two spectra were compared using a $\chi^{2}$-minimization technique looking for any wavelength-dependent calibration distortions across each echelle order and across the wavelength range covered by all the echelle orders collectively. The overall velocity distortion was obtained by a linear fit of all the spectral regions and it was used to distortion-correct the corresponding quasar exposure.

\subsection{Supercalibration data}
\label{subsec:supercalidata}

\subsubsection{Data from 2013 and 2014}
\label{subsubsec:supercali2014}
The exposures taken in 2013 and in 2014 were recorded with an attached ThAr wavelength calibration. Since the problems of long-range distortions and their effect on $\mu$-variation analysis from H$_2$ spectra had been well identified and reported \mbox{\citep{Rahmani2013,Bagdonaite2014}}, supercalibration exposures targeting asteroids and solar twins were purposely recorded immediately after the science exposures with their attached ThAr calibrations. In view of their accessibility during observations of J1237+0647, supercalibration spectra of the Eunomia asteroid, as well a solar twin stars HD~76440, HD~147513, HD~097356 and HD~117860, were used. Only some exposures, which had to be aborted due to bad weather conditions, were not supported by a supercalibration. 

For each supercalibration spectrum, its long-range wavelength distortion was determined in terms of a slope parameter, which was found to range between 300 and \mbox{725 ms$^{-1}$} per \mbox{1000 \AA}. In particular, during each observing run the values of the distortion slopes showed an apparent stability, with a spread in the slope values of, at most, \mbox{$\sim$200 ms$^{-1}$} per \mbox{1000 \AA} in May 2014. These values, and in particular the positive sign, are commensurate with values found previously in such measurements \mbox{\citep{Rahmani2013,Whitmore2015}}. Results of the present analysis are plotted in \mbox{Fig. \ref{supercals}} and numerical values listed in \mbox{Table \ref{supercaldataset}}. Thereupon each individual J1237+0647 science exposure was corrected for its long-range distortion by introducing a counter-distortion based on these slopes, after which the exposures were rebinned and combined using \textsc{UVES\_popler}. Exposures without a directly attached supercalibration were corrected using the average value of all the distortion slopes from the supercalibrations taken during the same night, in view of the stability discussed above.

\subsubsection{Data from 2009}
\label{subsubsec:supercali2009}
The data for J1237$+$0647 taken in 2009, retrieved from the ESO archive, were not supported by an attached ThAr calibration, and they were therefore wavelength calibrated against a regular ThAr calibration recorded at the end of the night. No supercalibration spectra were recorded with the J1237+0647 science exposures. In the ESO archive some suitable asteroid observations were found as three exposures of the Ceres asteroid, recorded in March and April 2009 within one week of the quasar exposures, under program 080.C-0881(B) (PI Dumas). These exposures did not share the same telescope settings as the J1237+0647 observations. In particular they had different grating settings, covering only half of the H$_{2}$ window, as shown in Fig. \ref{supercals}, and they did not have an attached ThAr calibration. Unlike the quasar observations, these exposures were not taken with the slit perpendicular to the horizon and hence an atmospheric dispersion corrector was used not to lose the blue part of the flux due to the atmospheric dispersion. Such differences can affect the light path of Ceres within the telescope, resulting in a different one from the quasar light path. Since the causes of the long-range distortions are unclear \mbox{\citep{Whitmore2015}}, it is not possible to estimate the impact of those differences on the supercalibration process.

Analysis of the Ceres exposures yields two distortion slopes of \mbox{of $\sim 150$ ms$^{-1}$ per 1000 \AA}, while a third one shows a negative distortion \mbox{of $\sim -500$ ms$^{-1}$ per 1000 \AA}. The presence of two  values showing distortions with opposing signs for the same night is remarkable, in particular since these two Ceres spectra were calibrated using the same non-attached ThAr spectrum, and it prevents one to calculate a reliable average value of the distortion correction for these exposures. For these reasons, it was decided to not perform a counter-distortion analysis of the wavelength scale.

Nevertheless, a magnitude of the wavelength distortion can be estimated using the limiting values of the distortion slopes, given in Table \ref{supercaldataset}. Both limiting values of opposing signs were considered as the largest distortions that could have affected the UVES exposures of J1237+0647 in 2009. Therefore the full sub-spectrum formed from 2009 exposures was first counter-distorted using the limit values of the distortion slope and then was combined with the distortion-corrected spectrum formed from exposures taken in 2013 and 2014 to derive the constraints on $\Delta\mu/\mu$. The two values derived were \mbox{$\Delta\mu/\mu\vert_{\textrm{\small{low}}} = (-4.6 \pm 6.1_{\textrm{\small{stat}}}) \times 10^{-6}$} and \mbox{$\Delta\mu/\mu\vert_{\textrm{\small{up}}} = (0.1 \pm 6.1_{\textrm{\small{stat}}}) \times 10^{-6}$} for the lower and for the upper limit respectively. The spread in resulting values for \mbox{$\Delta\mu/\mu$} was interpreted in terms of a systematic uncertainty amounting to $2.4 \times 10^{-6}$.

\subsubsection{Other sources of systematics}
\label{subsec:systs}
\begin{figure}
\centering
\includegraphics[width=\columnwidth]{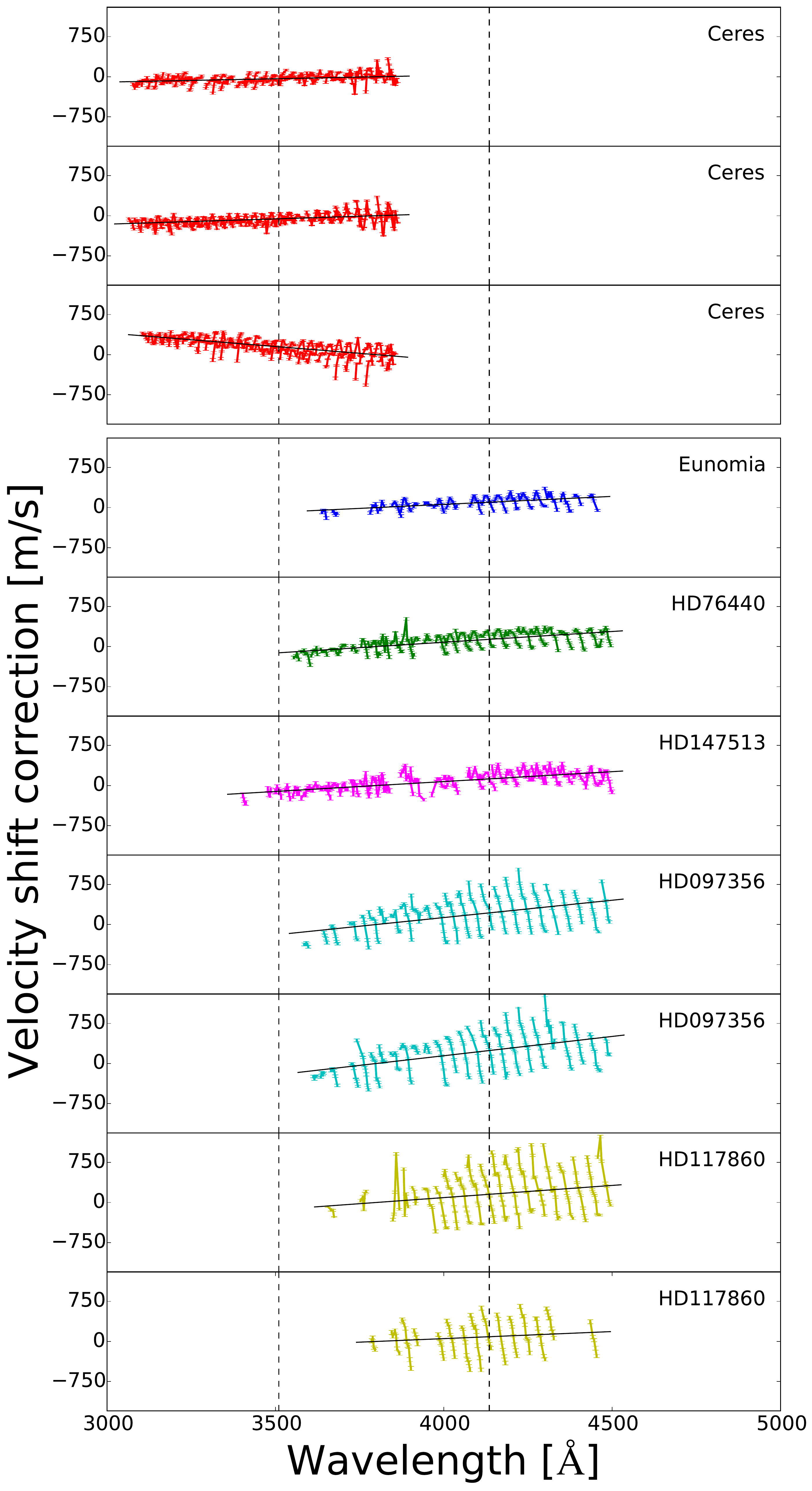}
\caption{Distortion maps of the UVES wavelength scale derived from asteroid and solar twin targets used for the supercalibration of science exposures of J1237+0647. For each supercalibration exposure, the plot shows the velocity shift required to align the ThAr-calibrated wavelength scale with that derived from the FTS solar spectrum. One observes both intra-order and long-range distortions of the wavelength scale. Only the slopes of the latter, whose values are listed in Table \ref{supercaldataset} with a one-to-one correspondence, are the physically relevant parameters used in the distortion analysis. They were determined by least-squares fits to the mean velocity shift correction from each echelle order as a function of wavelength. The two dashed vertical lines show the window in which the H$_{2}$ absorption lines toward J1237+0647 fall. The radial velocities are unimportant for the analysis of the slopes and have been removed here; the distortion maps have all been shifted to zero velocity shift at 3800 \AA. Top panel: three supercalibrations extracted from Ceres exposures taken in 2009 and retrieved from the ESO archive. Lower panel: seven supercalibrations extracted from asteroid and solar twin observations performed in 2013 and 2014. Eunomia, HD76440 and HD147513 were observed in visitor mode in May 2013 while the solar twins HD097356 and HD117860 were observed in service mode in 2014.}
\label{supercals}
\end{figure}

The supercalibrations allow to address the issues of both the intra-order and the long-range distortions that are known to affect UVES\citep{Molaro2008,Griest2010,Whitmore2010,Rahmani2013,Whitmore2015}. All the supercalibration targets used in this analysis are presented in Fig. \ref{supercals}. In each panel, the correction that needs to be applied to the ThAr wavelength is shown; note that each echelle order is characterized by several (typically 4-9) measurements separated \mbox{by $\sim$8 \AA}. The vertical spread in each echelle order reflects the magnitude of the intra-order distortions, while the slopes, obtained by least-squares fits to the mean velocity shift correction from each echelle order as a function of wavelength \mbox{\citep[as in][]{Whitmore2015}}, represents the long-range distortions.

The impact of the intra-order distortions is, typically, not expected to be dominant in the systematic error budget on the fiducial value of $\Delta\mu/\mu$, because the H$_{2}$ and HD transitions are spread across multiple orders \mbox{\citep{Malec2010,King2011,Bagdonaite2014}}. Nevertheless, the supercalibrations taken in 2014 show intra-order distortions that are \mbox{$\sim$ three} times larger than the distortions in the exposures taken in 2009 and 2013. An estimation of the impact of the intra-order distortions on the systematic error on $\Delta\mu/\mu$ is made using \mbox{$\delta(\Delta\mu/\mu) = [(\Delta v/c)/\sqrt{N}]/\Delta K_{i}$}, where \mbox{$\Delta v/c$} is the mean amplitude of the intra-order distortions, \mbox{$N=137$} is the number of H$_{2}$ and HD transitions detected and \mbox{$\Delta K_{i}=0.065$} is the spread in the sensitivity coefficients. 

The supercalibration exposures return a mean velocity shift of \mbox{$\sim$ 25 ms$^{-1}$} for 2009 and 2013 and \mbox{$\sim$ 60 ms$^{-1}$} for 2014. The shifts translate into systematic uncertainties on $\Delta\mu/\mu$ of \mbox{$1.2\times10^{-6}$} and \mbox{$3.2\times10^{-6}$} respectively. The average value of $\delta(\Delta\mu/\mu)=2.2\times10^{-6}$ is taken as the systematic uncertainty on the $\mu$-variation due to the intra-order distortions. However, this estimation does not take into account various factors like the different sensitivity of the transitions, where they fall with respect to the order centre and the variation of the S/N across the H$_{2}$ window. As a consequence, this value should be considered as an upper limit \mbox{\citep{Malec2010}}.

Long-range and intra-order distortions are not the only contributors to the systematic error budget. In addition, other effects, like the spectral redispersion, the presence of exposures without any ThAr attached calibration and the calibration residuals, contribute to the total systematic error budget. Previous studies addressed these possible error sources, finding that the absence of attached ThAr calibrations introduces an error of  $\sim 0.7 \times 10^{-6}$ \mbox{\citep{Bagdonaite2014}}, the calibration residuals have an effect of, at most, \mbox{$\sim2.0\times10^{-6}$} \mbox{\citep{Murphy2007,Bagdonaite2014}}, and the spectral redispersion can contribute up to \mbox{$\sim 1.4\times10^{-6}$} for a pixel size of \mbox{$\sim 2.5$\kms} \mbox{\citep{King2011}}. Adding in quadrature all these contributions to the systematic error budget yields a systematic error on the fiducial value of $\Delta\mu/\mu$ of $\sim 4.0 \times 10^{-6}$.

\subsection{Constraining $\Delta\mu/\mu$ from the distortion-corrected spectrum}
\label{subsec:velshiftdm}
After having distortion-corrected the exposures taken in 2013 and 2014, an updated value of \mbox{$\Delta\mu/\mu\vert_{\textrm{\small{2014}}} = (1.2 \pm 11.8_{\textrm{\small{stat}}}) \times 10^{-6}$} was derived from the sub-spectrum formed from those exposures only. This value, which agrees within 1$\sigma$ with the constraint derived only from exposures from 2009, is represented as a square in Fig. \ref{tests}.

The `final spectrum', obtained combining the distortion corrected exposures together with the uncorrected exposures from 2009, delivers a fiducial constraint on the proton-to-electron mass ratio of \mbox{$\Delta\mu/\mu = (-5.4 \pm 6.3_{\textrm{\small{stat}}} \pm 4.0_{\textrm{\small{syst}}}) \times 10^{-6}$}, which is the final result of the present study.
\begin{table*}
\centering
\caption{Observations with VLT/UVES of the supercalibration targets used in this work. Data from program 080.C-0881(B)  were retrieved from the ESO archive and have only the regular ThAr calibration taken at the end of the night. The slit width was \mbox{1.0 arcsec} for all exposures. The uncertainty on the slopes is $\sim$ 30 ms$^{-1}$ per 1000 \AA.}
\label{supercaldataset}
\begin{tabular}{ll*{3}{c}}
\hline
Target & Program ID & Date & Grating [nm] & Distortion slope \\
 & & & & [ms$^{-1}$ per 1000 \AA] \\
\hline
\hline
Ceres & 080.C-0881(B) & 23-03-2009 & 346 & 120 \\
 & & 01-04-2009 & 346 & 200 \\
 & & 01-04-2009 & 346 & $-510$ \\
Eunomia & 091.A-0124(A) & 15-05-2013 & 390+580 & 300 \\
HD76440 & & 15-05-2013 & 390+580 & 400 \\
HD147513 & & 16-05-2013 & 390+580 & 370 \\
HD097356 & 093.A-0373(A) & 23-03-2014 & 390+580 & 650 \\
 & & 03-04-2014 & 390+580 & 720 \\
HD117860 & & 28-05-2014 & 390+580 & 460 \\
 & & 31-05-2014 & 390+580 & 260 \\
\hline
\end{tabular}
\end{table*}

\section{Conclusion}
\label{sec:conclusions}
In this work, the analysis of molecular hydrogen absorption in the system at redshift $z=2.69$, in the line of sight toward quasar J1237+0647, is presented in order to constrain a possible variation of the proton-to-electron mass ratio $\mu$. 137 H$_{2}$ and HD transitions, found in three distinct velocity features associated with the DLA, some even exhibiting further velocity fine-structure, were analyzed. The large number of absorption features considered in the dataset includes partially overlapped features, due to the complex velocity structure of the system, as well as strongly saturated lines. This was possible because of the comprehensive fitting method used in this analysis. Intra-order and long-range distortions, which are known to affect the UVES spectra, were taken into account by applying the supercalibration technique presented by \mbox{\cite{Whitmore2015}} to the exposures taken in 2013 and 2014. For exposures taken in 2009, and retrieved from the ESO archive, the impact of such distortions was estimated using asteroid observations taken within one week of the quasar ones, and it was translated into a contribution to the systematic error budget. The resulting value of a constraint on a varying proton-to-electron mass ratio is \mbox{$\Delta\mu/\mu = (-5.4 \pm 6.3_{\textrm{\small{stat}}} \pm 4.0_{\textrm{\small{syst}}}) \times 10^{-6}$}. 

This constraint can be improved using the upcoming new generation of high resolution spectrographs, like the Echelle SPectrograph for Rocky Exoplanet and Stable Spectroscopic Observations \mbox{\citep[ESPRESSO,][]{Pepe2010}}. The spectral resolution achieved in this work allows to resolve components with Doppler broadening parameters \mbox{\emph{b} $\sim2.5$\kms},  while the high resolution of the new instruments, roughly three times higher than used here, will allow to resolve components with Doppler broadening parameters of the order \mbox{of $\sim$ 1.0\kms}. This may help further resolve finer structure within individual spectral features and hence result in a more precise $\Delta\mu/\mu$ constraint from \mbox{J1237$+$0647}. Moreover, the new instrument with fiber-fed and frequency-comb support should not suffer from the long-range wavelength distortions  that are affecting UVES, and should lead to smaller systematic uncertainties in $\mu$-variation analyses.

\begin{figure}
\centering
\includegraphics[width=1\columnwidth]{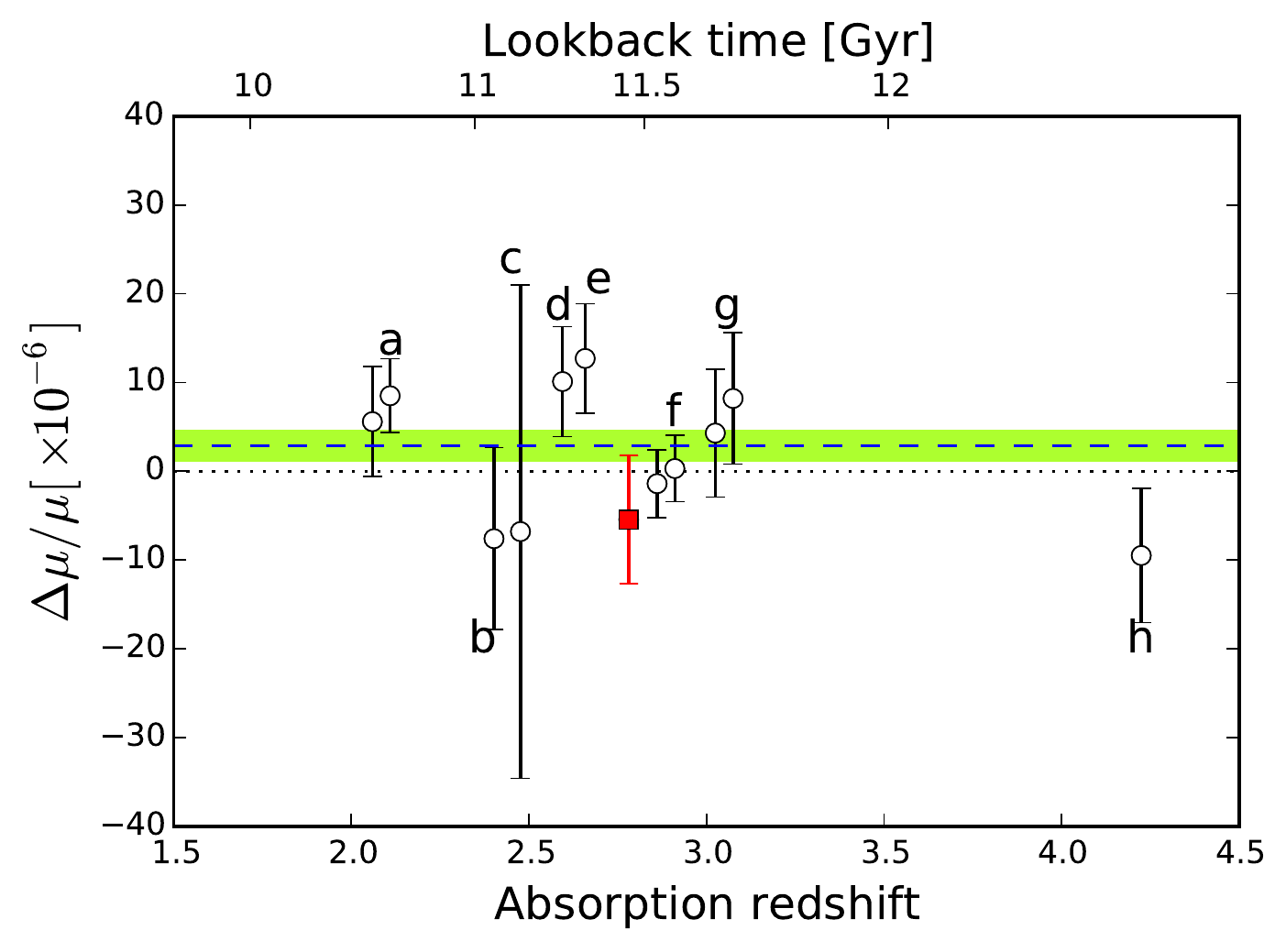}
\caption{Overview of results from investigations on a varying proton-to-electron mass ratio using molecular hydrogen absorbing systems. The result on \mbox{J1237+0647} presented in this work is indicated with a (red) square. The circles show results from previous analyses on systems: a) \mbox{J2123--0050}~\citep{Malec2010,Weerdenburg2011}, b) \mbox{HE0027--1836}~\citep{Rahmani2013}, c) \mbox{Q2348--011}~\citep{Bagdonaite2012}, d) \mbox{Q0405--443}~\citep{King2008}, e) \mbox{B0642--5038}~\citep{Bagdonaite2014}, f) \mbox{Q0528--250}~\citep{King2008,King2011}, g) \mbox{Q0347--383}~\citep{King2008,Wendt2012} and h) \mbox{Q1443+272}~\citep{Bagdonaite2015}. Note that multiple values for the constraint on $\Delta\mu/\mu$ were derived from systems J2123--0050 (labelled as `a'), Q0528--250 (labelled as `f') and Q0347--383 (labelled as `g'). To avoid overlaps between observations of the same system, their points are presented with an offset of +0.05 on the \emph{z}-scale. The dotted line represents the zero level, while the dashed line shows the weighted mean of all the $\Delta\mu/\mu$ values and the shaded area shows its $1\sigma$ boundaries.
}
\label{summary}
\end{figure}

In Fig.~\ref{summary}  the result of the present work is compared with other constraints, taken from literature, on a varying $\mu$ obtained from previous investigations on other H$_2$ absorbers at medium-to-high redshifts. The weighted mean, obtained considering, where possible, both the statistical and the systematic errors, resulting from this larger set is \mbox{$\Delta\mu/\mu=(2.9 \pm 1.7)\times10^{-6}$}. This is consistent with no variation of $\mu$ over a cosmological time-scale at a level of $\sim$10$^{-5}$ for a look-back time of \mbox{$\sim$10.5--12.5 Gyrs}.

•

\section*{Acknowledgments}
The authors thank the Netherlands Foundation for Fundamental Research of Matter (FOM) for financial support. MTM thanks the Australian Research Council for \textsl{Discovery Project} grant DP110100866 which supported this work. The work is based on observations with the ESO Very Large Telescope at Paranal (Chile).

\bibliographystyle{mn2e}
\bibliography{bibliography}

\appendix
\section{Absorption model}
\label{app:model}
\begin{figure*}
\centering
\includegraphics[width=2\columnwidth]{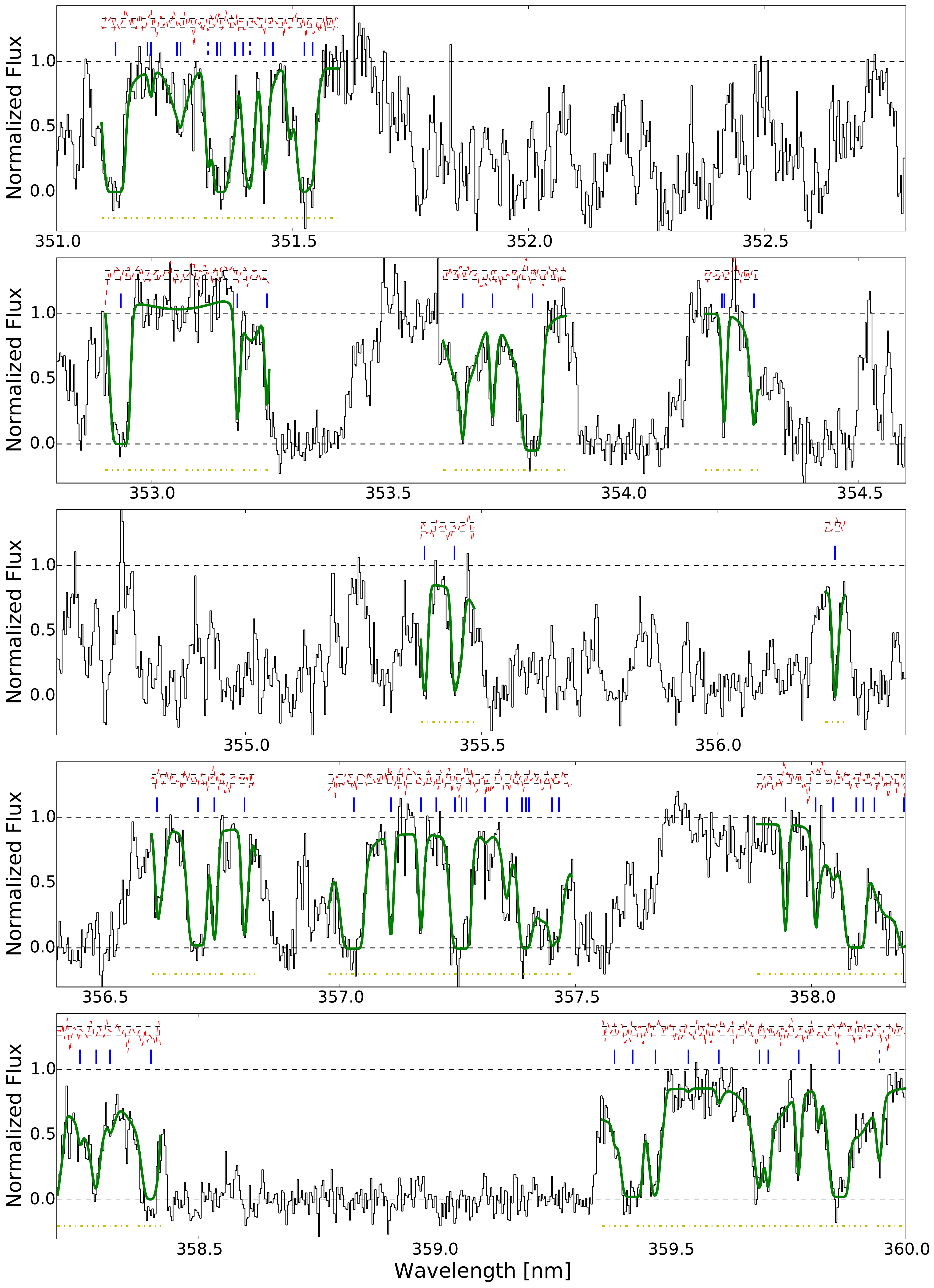}
\caption{Spectrum of quasar J1237+0647 (part 1 of 7). Regions containing the H$_{2}$ transitions used in this analysis are indicated by a green line of the fitted model. H$_{2}$ transitions are marked with a vertical solid line, HD transitions are marked with a dashed line and metal transitions are marked with a dotted line.}
\label{spectrum01}
\end{figure*}

\begin{figure*}
\centering
\includegraphics[width=2\columnwidth]{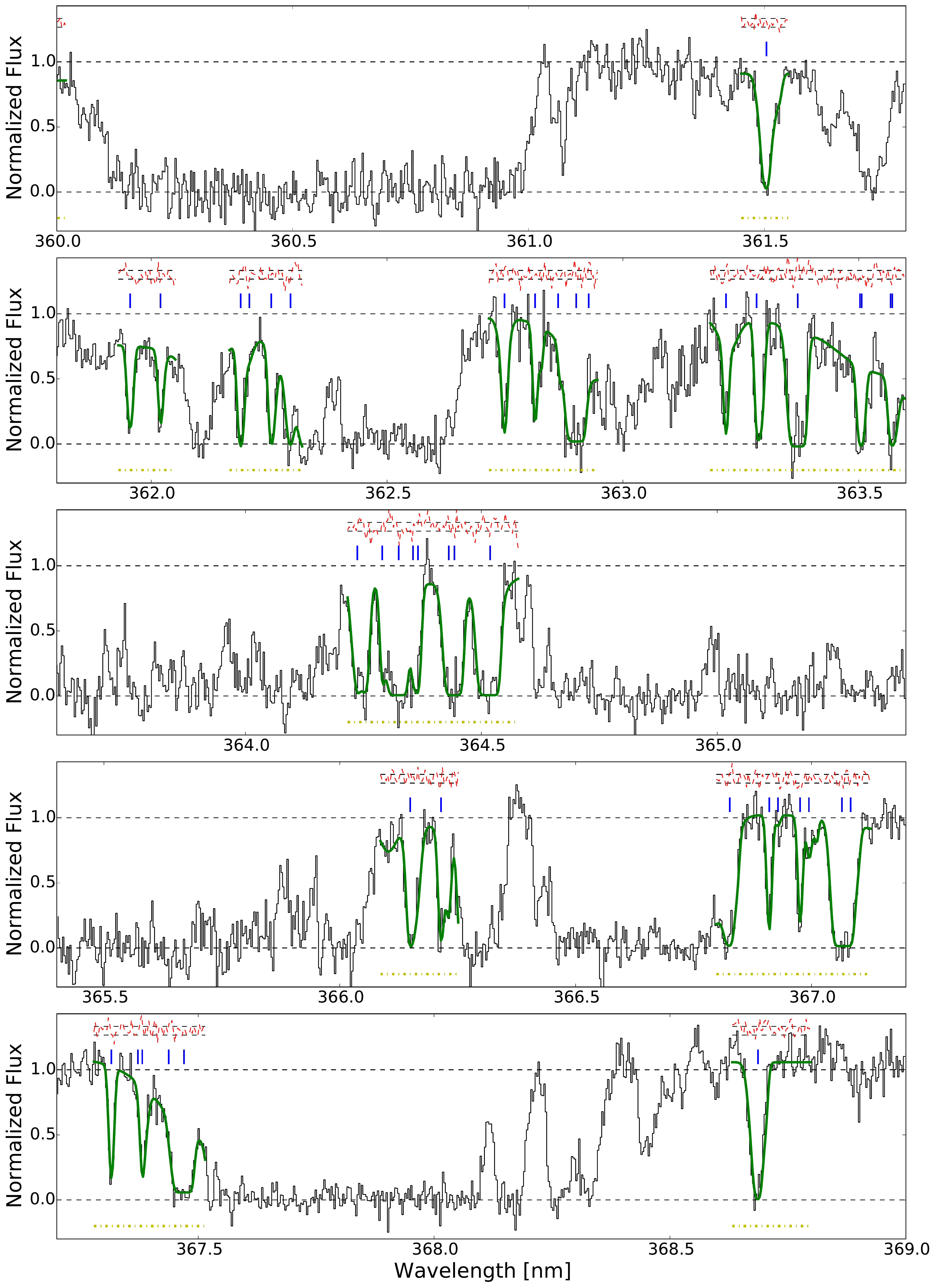}
\caption{Spectrum of quasar J1237+0647 (part 2 of 7), continued.}
\label{spectrum02}
\end{figure*}

\begin{figure*}
\centering
\includegraphics[width=2\columnwidth]{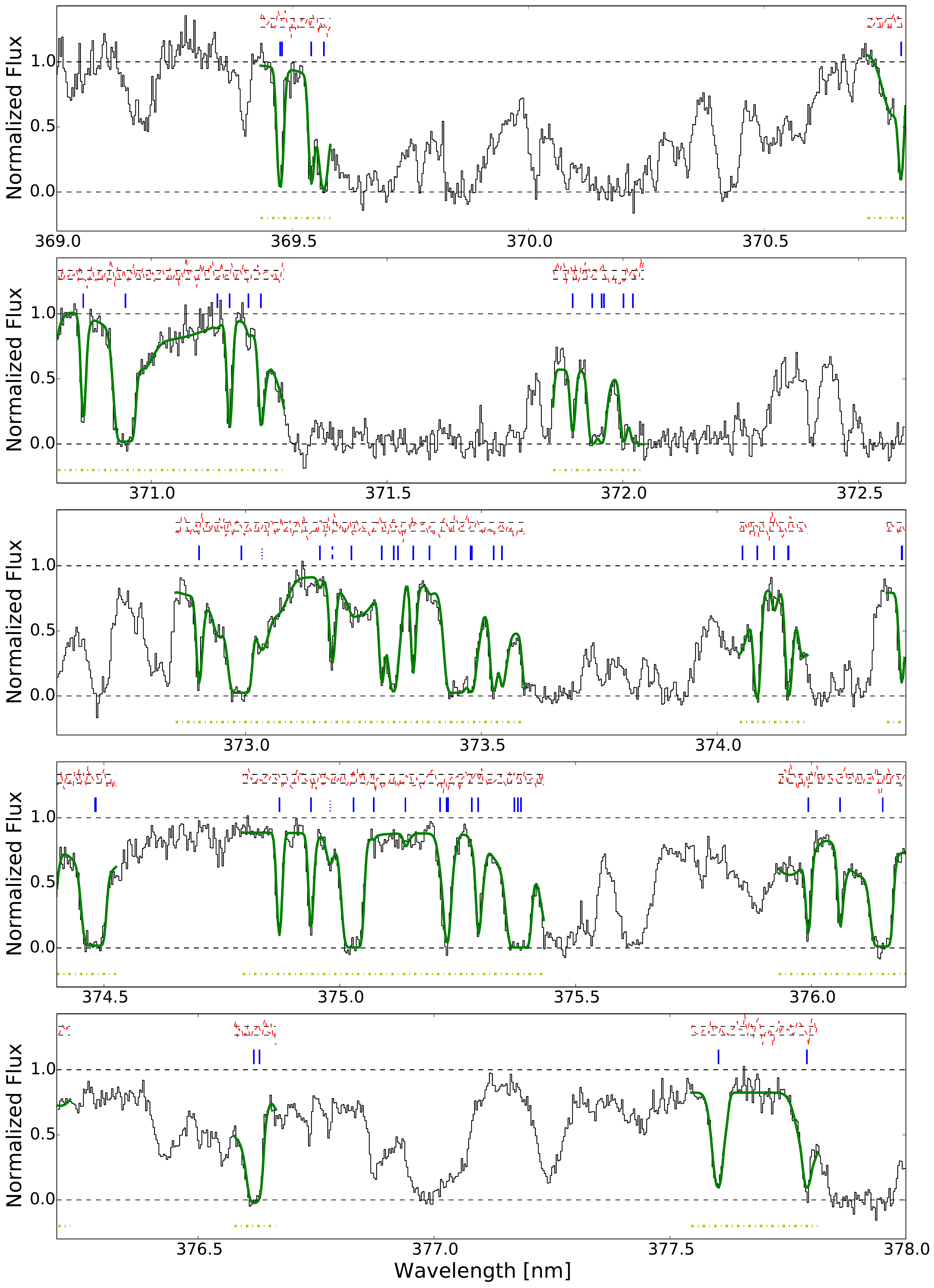}
\caption{Spectrum of quasar J1237+0647 (part 3 of 7), continued.}
\label{spectrum03}
\end{figure*}

\begin{figure*}
\centering
\includegraphics[width=2\columnwidth]{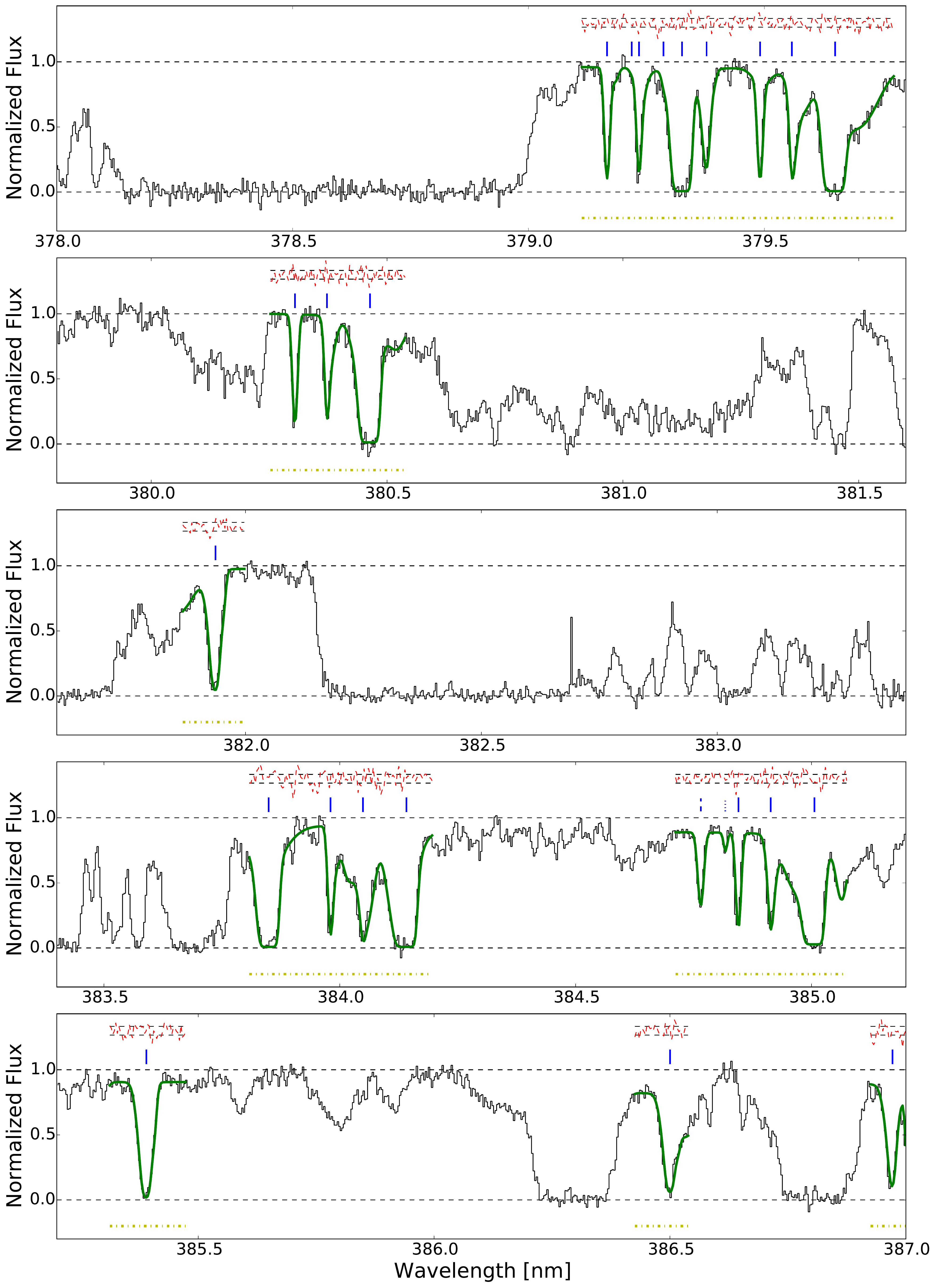}
\caption{Spectrum of quasar J1237+0647 (part 4 of 7), continued.}
\label{spectrum04}
\end{figure*}

\begin{figure*}
\centering
\includegraphics[width=2\columnwidth]{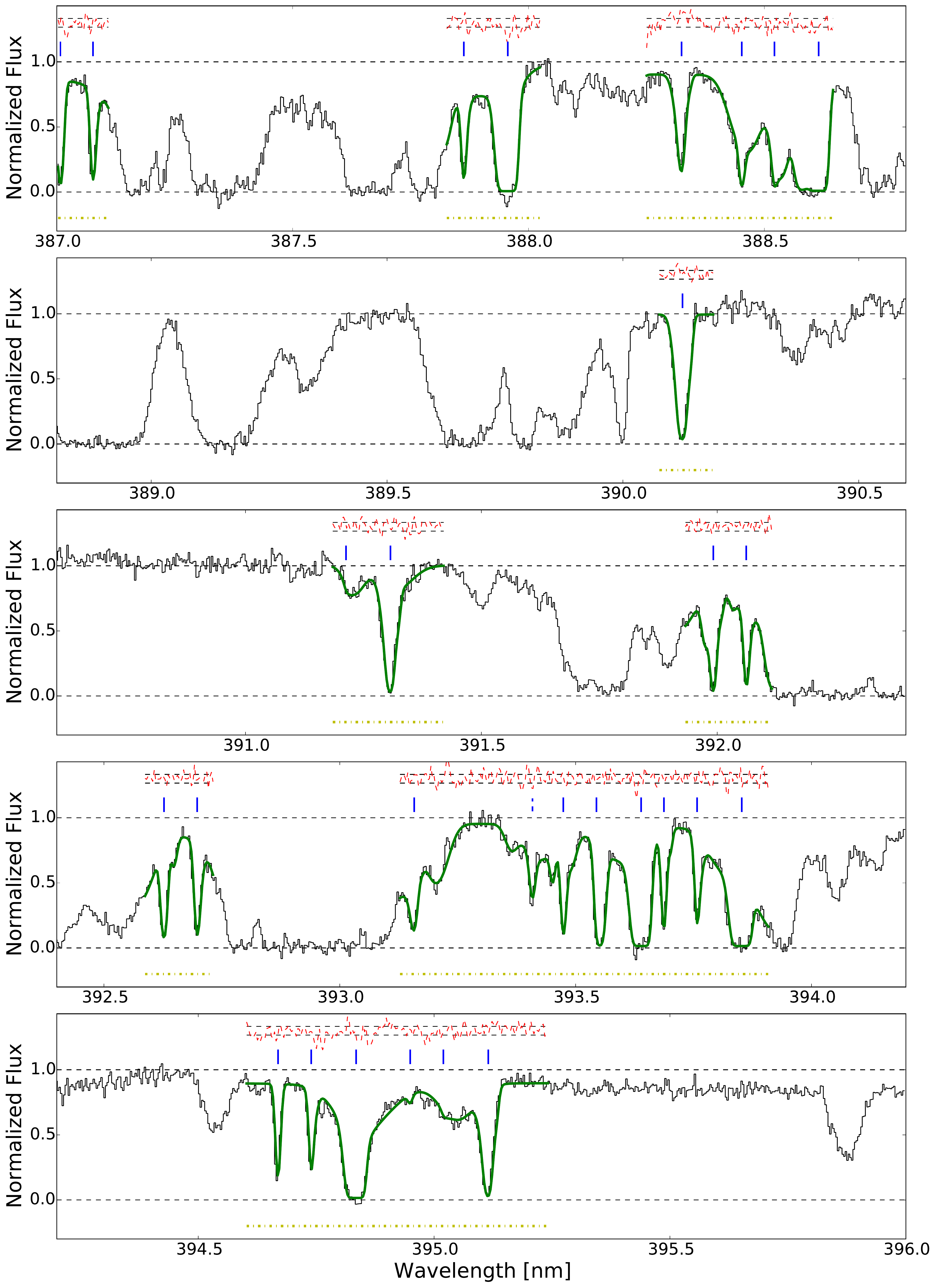}
\caption{Spectrum of quasar J1237+0647 (part 5 of 7), continued.}
\label{spectrum05}
\end{figure*}

\begin{figure*}
\centering
\includegraphics[width=2\columnwidth]{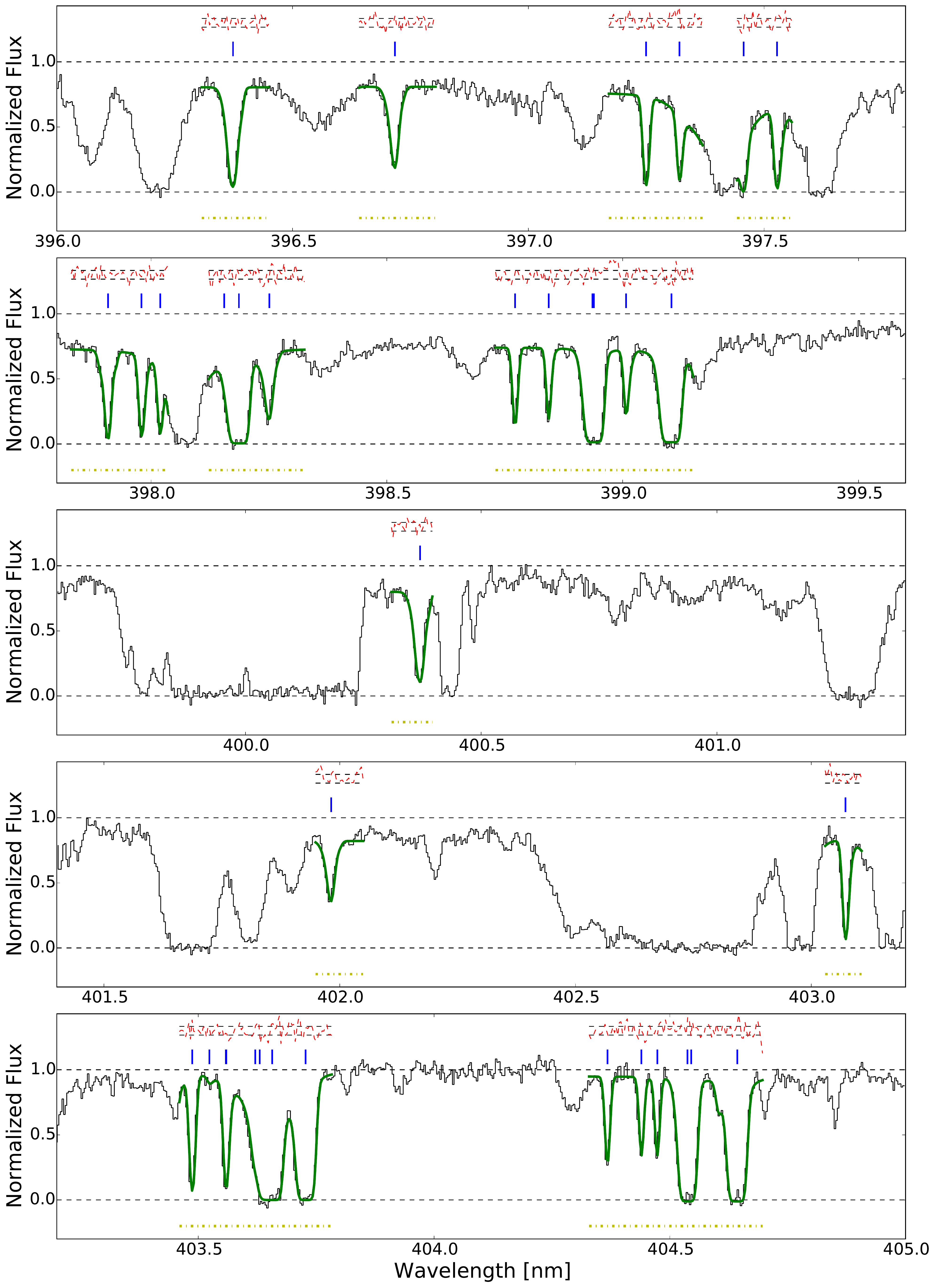}
\caption{Spectrum of quasar J1237+0647 (part 6 of 7), continued.}
\label{spectrum06}
\end{figure*}

\begin{figure*}
\centering
\includegraphics[width=2\columnwidth]{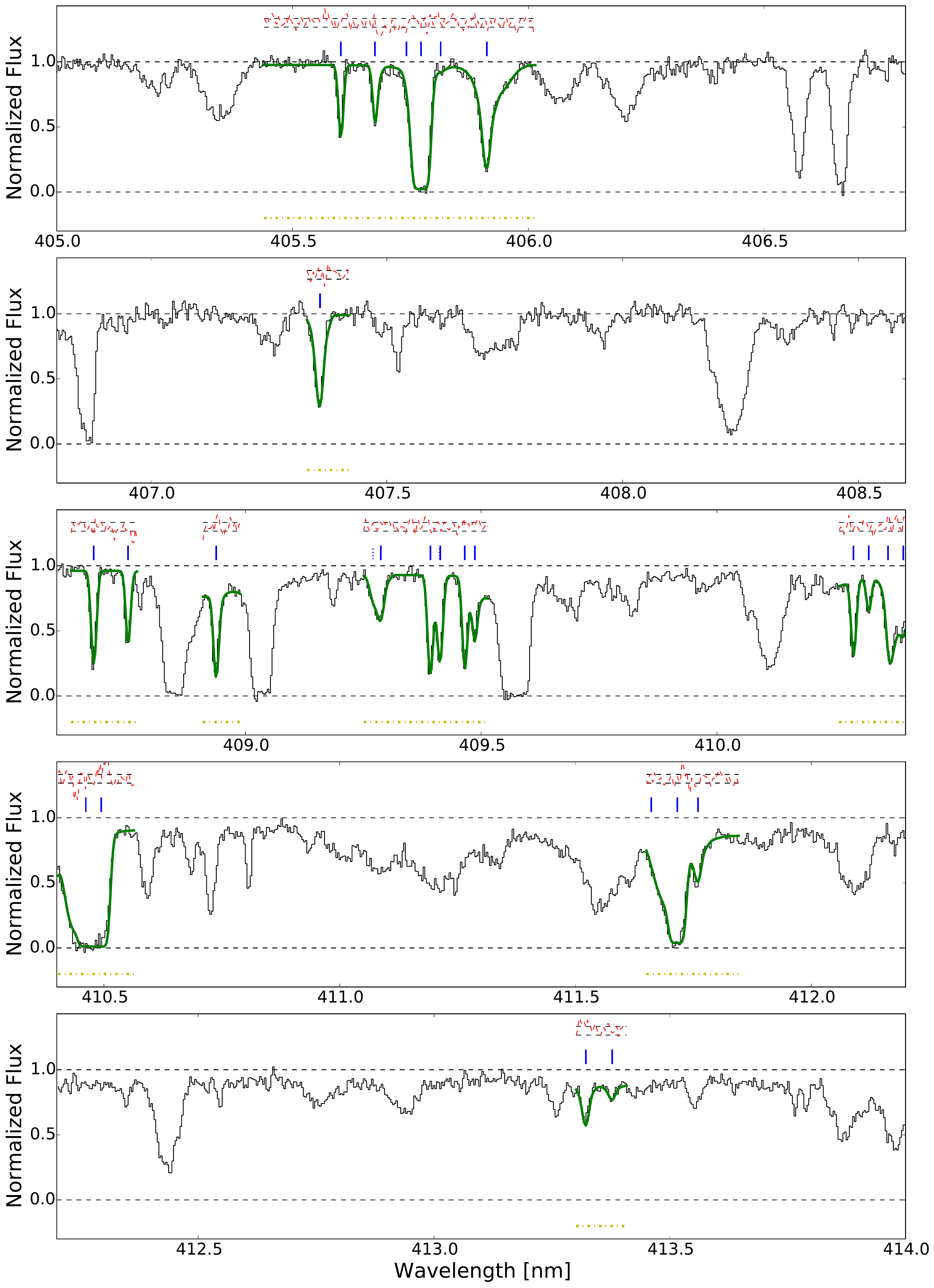}
\caption{Spectrum of quasar J1237+0647 (part 7 of 7), continued.}
\label{spectrum07}
\end{figure*}

Figures \ref{spectrum01}--\ref{spectrum07} show the H$_{2}$ window in the spectrum of the absorbing system at \mbox{$z \approx 2.69$} toward quasar \mbox{J1237+0647} along with the absorption model.

•
•
•
•
•
•
•
\bsp

\label{lastpage}

\end{document}